# Photoionization of atoms stuffed inside a two-shell fullerene

**M. Ya. Amusia**[1,2], **L. V. Chernysheva**[1], **E. Z. Liverts**[2]

[1] A. F. Ioffe Physical-Technical Institute, 194021 St. Petersburg, Russia
[2]Racah Institute of Physics, Hebrew University, 91904 Jerusalem, Israel

## Abstract

The photoionization of a two-shell endohedral $A@C_{N1}@C_{N2}$ is considered. Formulas are presented for cross-sections and angular anisotropy parameters, both dipole and non-dipole. The effect of the fullerenes shell upon photoelectron from atom A is taken into account substituting the action of the fullerene by two zero-thickness "bubble potential". The fullerenes shells polarization is included assuming that the radius of the outer shell $R_2$ is much bigger than the inner $R_1$ and both much exceeds the atomic radius $r$. This permits to express the effect via $C_{N1,\ N2}$ polarizabilities, which are connected to their photoionization cross-sections.

The interaction between shells $C_{N1}$ and $C_{N2}$ is taken into account in the Random Phase Approximation (RPA). The effect of photoelectron scattering by both "bubble potentials" is included in the lowest order and in the RPA frame.

As concrete examples, two endohedrals $Ar@C_{60}@C_{240}$ and $Xe@C_{60}@C_{240}$ are considered. In the first case we consider 3p and 3s, while in the second – 5p, 5s and 4d subshells. A whole variety of new peculiarities are found that deserve experimental verification.

## 1. Introduction

A lot of attention is given in recent years to photoionization of endohedral atoms [1-4]. These are objects consisting of a fullerene $C_N$ and an atom A, stuffed inside, $A@C_N$. The attention is concentrated on the modification of this atom photoionization characteristic. Indeed, the difference of them as compared to that of isolated atom A gives information on the fullerenes structure. In a sense, the inner atom A in $A@C_N$ serves as a "lamp" that shines "light" in the form of photoelectron waves that "illuminates" the fullerene $C_N$ from the inside.

A number of specific features were predicted in photoionization of $A@C_N$ that makes it different from photoionization of an isolated atom A itself. Most prominent features are the so – called confinement [5] and Giant endohedral [6] resonances that are consequences of two prominent effects – the reflection of the photoelectrons by the fullerenes shell and modification of the incoming photon beam due to $C_N$ polarization [7, 8]. In spite of prominent theoretical efforts, it is quite a few experimental investigations in this area [9, 10].

As usual, what is easier to calculate is more difficult to measure and vice versa. As an object of calculations the almost ideally spherically symmetric fullerene $C_{60}$ is usually considered, while experiment is done for deformed $C_{80}$ and $C_{82}$. As inner atoms noble gases, hydrogen and alkali atoms are main objects of calculations, whereas in experiments Ce, Pr and their ions are considered. We do believe, however, that this process of mutual attempts will converge in not too distant future, demonstrating whether the developed calculation approaches are accurate enough.

Meanwhile, it appeared that fullerenes could be two-shell structures [11, 12]. It seems, therefore, timely to consider an endohedral of such type that we denote as $A@C_{N1}@C_{N2}$. In this paper we will study such an object using to describe reflection a rather simple approach that substitute both shells by infinitely thin potential layers. We will call it two-bubble potential. How to take into account a single bubble potential is described at length in e.g. [13]. Here we will present formulas for the two-bubble potential. We will assume that since the radius of the outer bubble is considerably smaller than the inner one, they are not affecting each other. This is why for experimentally known radiuses $R_1$, $R_2$ and corresponding number of carbon atoms in it, $N_1$, $N_2$, respectively we obtain the potential strength of both of them $V_1$ , $V_2$, using experimentally known electron affinities $I_1$, $I_2$ of fullerenes $C_{N1}$ and $C_{N2}$, respectively just as it was done for an ordinary bubble potential [13].

As it was already mentioned in a number of places (e.g. [14]), the “bubble” (or, more solemnly, “orange – skin”) potential is valid when the photoelectron’s wave-length is much bigger than the fullerenes thickness. In numbers, it means photoelectron’s energy up to 2-3 *Ry*. As a concrete example, we will consider two shells, with $N_1$=60 and $N_2$=240

Assuming that the fullerenes radius is much bigger than the atomic one, the effect of fullerenes polarization upon the incoming photon beam can be expressed via fullerenes polarizability [7]. The latter via dispersion relation is connected to the experimentally measurable fullerenes photoionization cross-section [15]. We will assume for simplicity that the outer fullerenes radius is much bigger than the inners one. In this approximation we will take into account not only the fullerenes action upon the atom’s photoionization, but mutual influence of both fullerenes shells as well. Since the photoionization cross-section of $C_{240}$ we will use rather approximate scaling to estimate the polarizability of this object.

## 2. Photoelectron scattering by two-bubble potential

The two-bubble potential is of the form

$$V(r) = -V_1\delta(r-R_1) - V_2\delta(r-R_2) \tag{1}$$

so that the equation for a photoelectron with the angular momentum $l$ and energy $E$ moving in the atomic potential $U(r)$ is of the form

$$\frac{1}{2}\left[\chi''_{kl} - \frac{l(l+1)}{r^2}\chi_{kl}\right] + \left[V_1\delta(r-R_1) + V_2\delta(r-R_2) - U(r) + E\right]\chi_{kl} = 0, \tag{2}$$

has solutions:

$$\chi_{kl}(r) = \begin{cases} F_l(k)u_{kl}(r) & (r \le R_1) \\ C_1u_{kl}(r) + C_2\upsilon_{kl}(r) & (R_1 < r \le R_2) \\ u_{kl}(r)\cos\delta_l - \upsilon_{kl}(r)\sin\delta_l & (R_2 < r) \end{cases}. \tag{3}$$

Here $k = \sqrt{2E}$ .

Let us introduce the following notations:

$$
\begin{aligned}
&F_l(k) \equiv F, \\
&u_{kl}(R_1) \equiv u_1, \\
&u_{kl}(R_2) \equiv u_2, \\
&\upsilon_{kl}(R_1) \equiv \upsilon_1, \\
&\upsilon_{kl}(R_2) \equiv \upsilon_2.
\end{aligned}
\tag{4}
$$

Then the condition of the wave function continuity (4) at $r = R_1$ acquires the form:

$$Fu_1 = C_1 u_1 + C_2 \upsilon_1 \tag{5}$$

Integration of equation (2) near the point $r = R_1$ gives:

$$Fu_1' = C_1 u_1' + C_2 \upsilon_1' + 2V_1 F u_1 . \tag{6}$$

Multiplying (5) and (6) by $u_1'$ and $u_1$, respectively, and then subtracting one equation from another, the following relation is obtained:

$$C_2 (u_1 \upsilon_1' - \upsilon_1 u_1') = -2V_1 F u_1^2 . \tag{7}$$

By inserting here the so-called Wronskian relation

$$u_{kl}(r)\upsilon_{kl}'(r) - \upsilon_{kl}(r) u_{kl}'(r) = k , \tag{8}$$

it is obtained:

$$C_2 k = -2V_1 F u_1^2 . \tag{9}$$

Excluding $F$ from (5) and (9), the first equation, connecting the coefficients $C_1$ and $C_2$ is obtained:

$$2V_1 u_1^2 C_1 = -(k + 2V_1 u_1 \upsilon_1) C_2 . \tag{10}$$

Let us consider now the point $r = R_2$ from the outside of the spherical layer. From the continuity conditions for the wave function (3) at this point, one obtains:

$$(C_1 u_2 + C_2 \upsilon_2) = u_2 \cos\delta_l - \upsilon_2 \sin\delta_l . \tag{11}$$

Integration of (2) near $r = R_2$ gives:

$$-2V_2 (C_1 u_2 + C_2 \upsilon_2) = u_2' \cos\delta_l - \upsilon_2' \sin\delta_l - C_1 u_2' - C_2 \upsilon_2' . \tag{12}$$

Multiplying (11) and (12) by $u_2'$ and $u_2$, respectively, and then subtracting the first equation from the second one, the following relation is obtained:

$$\sin \delta_l = -C_2 + \frac{2V_2 u_2 (C_1 u_2 + C_2 \upsilon_2)}{k}. \tag{13}$$

If one multiplies (11) and (12) instead by $\upsilon_2'$ and $\upsilon_2$, respectively, and then subtracts the first equation from the second one, another relation is obtained:

$$\cos \delta_l = C_1 + \frac{2V_2 \upsilon_2 (C_1 u_2 + C_2 \upsilon_2)}{k}. \tag{14}$$

In the derivation of (13) and (14), we used the equation (8).

Using the well-known relations for the trigonometric functions, the second relation, connecting the coefficients $C_1$ and $C_2$ is obtained:

$$[kC_1 + 2V_2 \upsilon_2 (C_1 u_2 + C_2 \upsilon_2)]^2 + [kC_2 - 2V_2 u_2 (C_1 u_2 + C_2 \upsilon_2)]^2 = k^2. \tag{15}$$

At last, solving the system of equations (10) and (15) for coefficients $C_1$ and $C_2$ we obtain:

$$C_1 = \frac{k(k + 2V_1 u_1 \upsilon_1)}{\sqrt{\zeta}}, \quad C_2 = -\frac{2kV_1 u_1^2}{\sqrt{\zeta}}, \tag{16}$$

where

$$\begin{aligned} \zeta = {} & k^4 + 4k^3 (V_1 u_1 \upsilon_1 + V_2 u_2 \upsilon_2) + 16 V_1^2 V_2^2 u_1^2 (u_2 \upsilon_1 - u_1 \upsilon_2)^2 (u_2^2 + \upsilon_2^2) - \\ & 16kV_1 V_2 u_1 (u_1 \upsilon_2 - u_2 \upsilon_1) \left[ V_1 u_1 (u_1 u_2 + \upsilon_1 \upsilon_2) + V_2 u_2 (u_2^2 + \upsilon_2^2) \right] + \\ & 4k^2 \left\{ V_1^2 u_1^2 (u_1^2 + \upsilon_1^2) + V_2^2 u_2^2 (u_2^2 + \upsilon_2^2) + 2V_1 V_2 u_1 \left[ 2u_2 \upsilon_1 \upsilon_2 + \upsilon_1 (u_2^2 - \upsilon_2^2) \right] \right\}. \end{aligned} \tag{17}$$

Inserting expressions (16) for $C_1$ and $C_2$ into (13), we obtain:

$$\sin \delta_l = \frac{2}{\sqrt{\zeta}} \left\{ kV_2 u_2^2 + V_1 u_1 \left[ k u_1 + 2V_2 u_2 (u_2 \upsilon_1 - u_1 \upsilon_2) \right] \right\}. \tag{18}$$

Removing the coefficient $C_2$ from (9) and (16), the following simple relation for the *reflection amplitude F* is obtained:

$$F \equiv F_l(k) = \frac{k^2}{\sqrt{\zeta}} = \frac{k \sin \delta_l}{2\left[ V_1 u_1^2 + V_2 u_2^2 - 2V_1 V_2 u_1 u_2 (u_1 \upsilon_2 - u_2 \upsilon_1) / k \right]}. \tag{19}$$

It is easy to verify that by putting $R_1 = R_2$ in (13), (14) and (17), the following relations are obtained

$$\begin{aligned} F_l(k) &= \frac{k \sin \delta_l}{2V_0 u_{kl}^2(R)} \\ \tan \delta_l &= \frac{u_{kl}^2(R)}{u_{kl}(R) \upsilon_{kl}(R) + k / 2V_0}, \end{aligned} \tag{20}$$

that coincides with corresponding relations for single-wall fullerene with $R_2 \to R$ and $(V_1 + V_2) \to V_0$ (see e.g. [16]).

## 3. Polarization effects from the two-shell fullerene

As it was already demonstrated quite a while ago, the following relation can account for the polarization of the fullerene under the action of the incoming photon beam [7]:

$$D_{AC}(\omega) \cong D_A(\omega)\left[1 - \frac{\alpha(\omega)}{R^3}\right], \qquad (21)$$

where $D_{AC}(\omega)$ is the endohedral atom A@$C_n$ photoionization dipole amplitude, $D_A(\omega)$ is the same for an isolated atom, $\alpha(\omega)$ is the fullerenes dipole polarizability, $R$ is its radius. This relation is derived under valid assumption that $R$ is much bigger than the atomic radius $r_A$.

It is convenient to represent the interaction of an incoming photon with an atom A stuffed inside a two-shell fullerene using diagrammatical approach, represented in the form suitable for atomic physics in e.g. [17]. The standard notations are used in the diagrams below: a horizontal dashed line stands for an incoming photon, vertically oriented wavy line represent the Coulomb inter-electron interaction, solid line with an arrow to the right (left) stands for electron (vacancy), respectively. The shadowed circle denotes the total amplitude of atom A photoionization with participation of both fullerenes shell.

The simplest way to take into account the fullerenes polarization is to consider the photoionization of A@$C_n$ in the frame of the Random Phase Approximation with Exchange (RPAE) that proved to be very successful in describing photoionization of multi-electron atoms [18]. If the atomic radius $r_A$ is much smaller than $R$, $R \gg r$, the photoionization amplitude $D_{AC}(\omega)$ of the atom A caged inside fullerene $C_n$ can be represented diagrammatically as [7]:

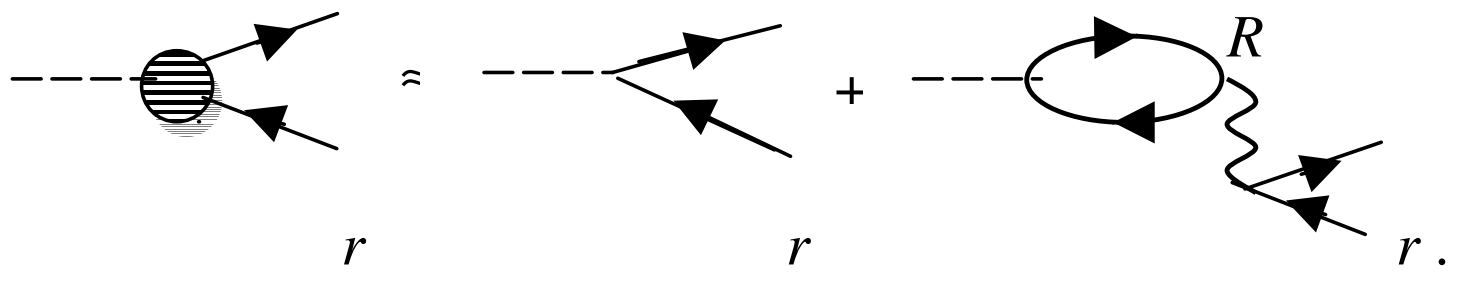

(22)

The first term in (22) represents the pure atomic photoionization amplitude while the second is a contribution to ionization of atom A via virtual excitations of the fullerenes shell. If $R \gg r$, equation (22) permits to go well beyond the lowest order term in inter-electron interaction by assuming that correlations between separately atomic and fullerenes electrons in RPAE or even out of its frame are included in the amplitude $D_A(\omega)$ and the fullerenes electron-vacancy loop. It can be demonstrated that higher order terms in atom-fullerene interaction is suppressed by the factor $\alpha^A(\omega)/R^3 \ll 1$, where $\alpha^A(\omega)$ is the A atom dipole polarizability.

For two-shell fullerene the corresponding expression is much more complex, since in principal the interaction between two group of electrons, belonging to fullerenes 1 and 2 must be taken into account. Under a reasonable qualitatively correct assumption, that holds at least for considered in this article $C_{60}$ and $C_{240}$, the following relation is correct $r \ll R_1 \ll R_2$. In this case the amplitude $D_{AC}(\omega)$ is presented by an infinite sequence of the following diagrams

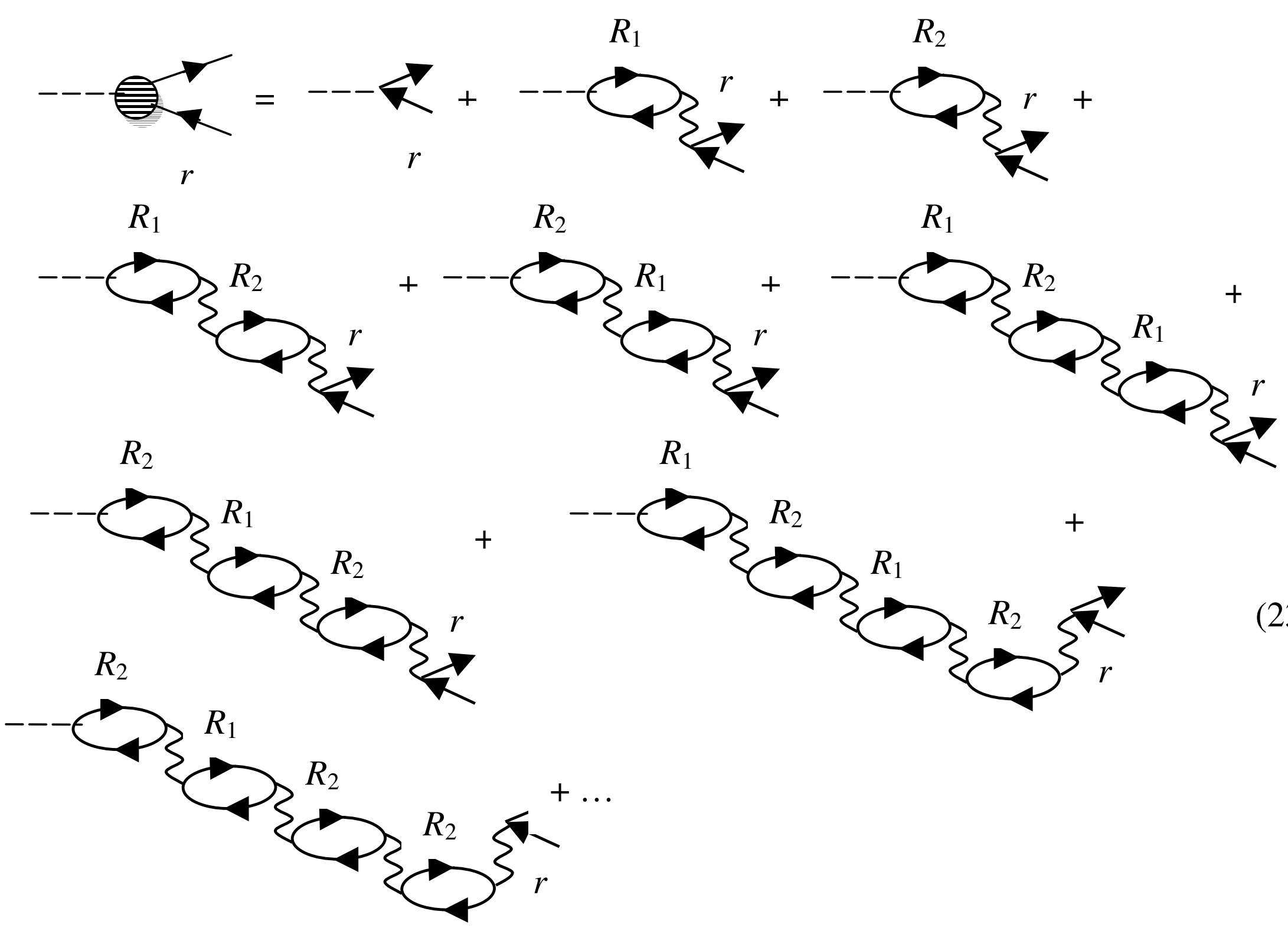

(23)

For simplicity of the drawings we have omitted so-called time-reverse diagrams.

The inequalities $r \ll R_1 \ll R_2$ permit to simplify the interactions between atomic and fullerene 1 and 2 electrons considerably, presenting it as $\vec{r}\vec{R}_1/R_1^2$, $\vec{r}\vec{R}_2/R_2^2$ and $\vec{R}_1\vec{R}_2/R_2^2$, respectively. The sequence can be summed, leading to appearance of the same denominator *den* to the contributions of all first and second order terms in the r.h.s. of (23)

$$den = 1 - \frac{\alpha_1\alpha_2}{R_2^6}, \tag{24}$$

where $\alpha_1 \equiv \alpha_1(\omega)$, $\alpha_2 \equiv \alpha_2(\omega)$ and $R_2$ are the dipole polarizability and radius of the outer fullerene, respectively. Finally, we arrive for $D_{AC}(\omega)$ to the following expression:

$$D_{AC}(\omega) \cong D_A(\omega)\left[1 - \left(\frac{\alpha_1}{R_1^3} + \frac{\alpha_2}{R_2^3}\right)\frac{1 - \dfrac{\alpha_1\alpha_2}{\alpha_1 R_2^3 + \alpha_2 R_1^3}\left(1 + \dfrac{R_1^3}{R_2^3}\right)}{1 - \dfrac{\alpha_1\alpha_2}{R_2^6}}\right] \equiv G_{12}(\omega)D_A(\omega), \tag{25}$$

where $G_{12}(\omega)$ is the *polarization amplitude factor* for two-shell fullerene.

It is seen that the correction due to simultaneous polarization of both fullerenes shells proportional to $\alpha_1\alpha_2$ considerably modifies a simple formula that would account only the sum of the shells action.

It is essential to have in mind that (25) can in principle take into account electron correlations beyond the RPAE frame. Namely, each of the polarizabilities, $\alpha_1$ or $\alpha_2$, can include even *all* correlations inside each fullerene, 1 or 2, respectively. It means that as polarizabilities, precise or best experimentally detected polarizabilities can be used.

### 4. Determination of polarizabilities

Polarizabilities $\alpha_1$ and $\alpha_2$ can be calculated or taken from experiment. Directly determined in experiment are only static polarizabilities, namely their values at $\omega \approx 0$. To obtain dynamic polarizability, calculations are needed. However, if photoionization cross-section $\sigma(\omega)$ of a considered object is measured, the dynamic polarizability can be derived using the following relations:

$$\begin{aligned} \operatorname{Im}\alpha(\omega) &= c\frac{\sigma(\omega)}{4\pi\omega} \\ \operatorname{Re}\alpha(\omega) &= \frac{c}{2\pi^2}\int_I^\infty \frac{\sigma(\omega')d\omega'}{\omega'^2-\omega^2}, \end{aligned} \qquad (26)$$

where $c$ is the speed of light and $I$ is the fullerenes ionization potential. Note that in the second relation in (26), so-called dispersion relation, it is assumed that the contribution of discrete excitations can be neglected. That is indeed confirmed by existing experimental data on fullerenes photoionization [19].

The results for polarizability obtained using (26) have to satisfy two constrains. The first is the static polarizability value that can be measured independently and has in principal to coincide with the following value:

$$\alpha(0) = \frac{c}{2\pi^2}\int_I^\infty \frac{\sigma(\omega)d\omega}{\omega^2}. \qquad (27)$$

The second constrain comes from the opposite limit of high frequencies of radiation $\omega \to \infty$, where by using the dipole sum rule the following relation is obtained

$$\operatorname{Re}\alpha_d(\omega \to \infty) = -\frac{1}{\omega^2}\frac{c}{2\pi^2}\int_I^\infty \sigma(\omega')d\omega' = -\frac{N}{\omega^2}, \qquad (28)$$

where $N$ is the total number of electrons in the considered fullerene.

In principal, not absolute but relative experimental data on $\sigma(\omega)$ are sufficient. This is because the relative values can be putted on the absolute scale using the sum rule:

$$(c/2\pi^2)\int_{I_o}^\infty \sigma(\omega)d\omega = N. \qquad (29)$$

While the experimental data on $\sigma(\omega)$ for $C_{60}$ are known, and $\alpha(\omega)$ can be reliably derived, this is not the case for $C_{240}$. This is why we will use a simple estimate scaling to obtain $\alpha(\omega)$ for this object (see below).

## 5. Derivation of parameters

To perform photoionization calculations we need to know the fullerenes potentials $V_1$ and $V_2$ that enter the equation (2). They can be determined using for each of the shells the same formula that is usually employed for a single-shell fullerene [13], so that

$$V_{0,1,2} = \frac{1}{2}\sqrt{2I_{0,1,2}}\left(1+\coth\sqrt{2I_{0,1,2}}R_{0,1,2}\right), \tag{30}$$

where $I$ is the ionization potential of the fullerene. By doing this we make a reasonable assumption that well separated shell are not essentially affecting each other so that a two-shell "onion" $C_{N_1,N_2}$ really consists of two fullerenes $C_{N_1}$ and $C_{N_2}$. We will take the electron affinities $I_1$ and $I_2$ for $C_{60}$ and $C_{240}$ from [20]. The values are $I_{60} = 0.195$ and $I_{240} = 0.280$. For completeness let us add $I_{540} = 0.386$. The radiuses of these fullerenes are also known, being equal to $R_{60} \cong 6.75$, $R_{240} \cong 13.5$ and $R_{60} \cong 19.8$ [1]. It is remarkable that for at least considered objects the ratio $\eta_{\rm n} \equiv N_{\rm n} / R_{\rm n}^2$ are almost the same $\eta_{60} = 1.317$, $\eta_{240} = 1.317$, $\eta_{540} = 1.38$. So, let us assume for $\eta_{\rm n}$ a universal value 1.32.

Analysis of the $C_{60}$ polarizability permitted to conclude that at least for static value $\alpha_{\rm C60} \approx 60\alpha_{\rm C}$. Since the electron density of all big enough fullerenes is the same, it seems natural to assume that $\alpha_{\rm C_n} \approx N_{\rm n}\alpha_{\rm C}$. In order to estimate the contribution of the second fullerenes shell, let us assume also that the same relations are valid not only for static but also for dynamic polarizabilities. Using the ratio $\eta_{\rm n} = 1.32$ and these assumptions, we obtain $\alpha_{\rm N}(\omega) / R_{\rm N}^3 \approx \alpha_{\rm C60}(\omega) / 45.5 R_{\rm N} \equiv \alpha_1(\omega) / 45.5 R_{\rm N}$. With the help of this relation we derive from (25)

$$D_{AC}(\omega) \approx D_A(\omega)\left[1 - 1.51\frac{\alpha_1}{R_1^3}\frac{1-0.380\alpha_1(\omega)/R_1^3}{1-0.0635[\alpha_1(\omega)/R_1^3]^2}\right] \equiv \bar{G}_{12}(\omega) D_A(\omega), \tag{31}$$

where $\bar{G}_{12}(\omega)$ is the approximate value of the polarization amplitude factor. The data on $\alpha_1(\omega)$ as well as on $S_1(\omega) = \left|\left(1-\alpha_1(\omega)/R_1^3\right)\right|^2$ are taken from [15].

## 6. Formulas for calculations

In this section we will present the formulas required to calculate the photoionization cross-sections and angular anisotropy parameters, both dipole and non-dipole for two-shell endohedral atom $\rm A@C_{N_1}@C_{N_2}$.

---

[1] All the values of electron affinity and radiuses are in atomic units that will be used throughout all this paper.

Let us start with the case, when the effects of fullerenes shells potential action upon photoelectrons of atom A is sufficient to take into account perturbative. In this case the following relation connects pure atomic partial photoionization cross-section $\sigma^{A}_{nl,kl'}(\omega)$ $(l'=l\pm 1)$ to the cross-section of the endohedral $A@C_{N_1}@C_{N_2}$

$$\sigma^{AC_{12}}_{nl,kl'}(\omega) = |F_{l'}(k)|^2 \, S_{12}(\omega)\sigma^{A}_{nl,kl'}(\omega) \qquad (32)$$

that is similar to that of the one-shell endohedral [14]. Here $S_{12}(\omega) = |G_{12}(\omega)|^2$ is the *polarization factor.*

The total photoionization cross-section is obtained from (32) by summing over $l'$:

$$\sigma^{AC_{12}}_{nl}(\omega) = \sum_{l'=l\pm 1} \sigma^{AC_{12}}_{nl,kl'}(\omega) \qquad (33)$$

The differential in angle $d\Omega$ cross-section of the photoelectron's emission under the action of non-polarized light is given for spherically-symmetric fullerenes shell, having the same center, by the following expression

$$\frac{d\sigma^{AC_{12}}_{nl}(\omega)}{d\Omega} = \frac{\sigma^{AC_{12}}_{nl}(\omega)}{4\pi}[1 - \frac{1}{2}\beta_{nl}(\omega)P_2(\cos\theta) + \kappa\gamma_{nl}(\omega)P_1(\cos\theta) + \kappa\eta_{nl}(\omega)P_3(\cos\theta)]. \quad (34)$$

Here $\kappa = \omega / c$, $c$ is the speed of light and $P_i(\cos\theta)$ are the Legendre polynomials.

The dipole angular anisotropy parameter $\beta_{nl}(\omega)$ is not affected by the fullerenes shell polarization amplitude and is given by

$$\beta_{nl}(\omega) = \frac{1}{(2l+1)\left[(l+1)F^2_{l+1}\tilde{D}^2_{l+1} + lF^2_{l-1}\tilde{D}^2_{l-1}\right]}[(l+1)(l+2)F^2_{l+1}\tilde{D}^2_{l+1}$$
$$+ l(l-1)F^2_{l-1}\tilde{D}^2_{l-1} - 6l(l+1)F_{l+1}F_{l-1}\tilde{D}_{l+1}\tilde{D}_{l-1}\cos(\tilde{\delta}_{l+1} - \tilde{\delta}_{l-1})] \quad . \qquad (35)$$

Here the dipole photoionization amplitudes $D_{l\pm1}(\omega)$ are complex numbers with module $\tilde{D}_{l\pm1}(\omega)$, their phases being determined by relation $D_{l\pm1}(\omega) \equiv \tilde{D}_{l\pm1}(\omega)\exp[i\Delta_{l\pm1}(k)]$. The phases $\tilde{\delta}_{l'}$ are determined by relation $\tilde{\delta}_{l'} = \delta_{l'} + \Delta_{l'}$. Here and below $F_{\nu_i}$ are the reflection factors, given by (19) with $\nu_i$ determined by photoelectron's linear momentum $k_{\nu_i} = \sqrt{2\varepsilon_{\nu_i}}$ and $l_{\nu_i}$. The corrections $G_{12}(\omega)$ are not entering $\beta_{nl}(\omega)$, since they are modifying the nominator and denominator in (35) similarly.

The situation with the non-dipole angular anisotropy parameters that are given by the expressions derived in [18] is different, since they include corrections due to dipole and quadrupole polarization amplitude $G^{d}_{12}(\omega)$ and $G^{q}_{12}(\omega)$:

$$\gamma_{nl}^{AC}(\omega)=\frac{3\tilde{G}_{12}^{q}(\omega)}{5\tilde{G}_{12}^{d}(\omega)\left[(l+1)F_{l+1}^{2}\tilde{D}_{l+1}^{2}+lF_{l-1}^{2}\tilde{D}_{l-1}^{2}\right]}\times$$
$$\times\left\{\frac{(l+1)F_{l+1}}{2l+3}[3(l+2)F_{l+2}\tilde{Q}_{l+2}\tilde{D}_{l+1}\cos\left(\tilde{\tilde{\delta}}_{l+2}-\tilde{\tilde{\delta}}_{l+1}\right)-lF_{l}\tilde{Q}_{l}\tilde{D}_{l+1}\cos\left(\tilde{\tilde{\delta}}_{l+2}-\tilde{\tilde{\delta}}_{l+1}\right)]-\right.$$
$$\left.-\frac{lF_{l-1}}{2l+1}\left[3(l-1)F_{l-2}\tilde{Q}_{l-2}\tilde{D}_{l-1}\cos\left(\tilde{\tilde{\delta}}_{l-2}-\tilde{\tilde{\delta}}_{l-1}\right)-(l+1)F_{l}\tilde{Q}_{l}\tilde{D}_{l-1}\cos\left(\tilde{\tilde{\delta}}_{l}-\tilde{\tilde{\delta}}_{l-1}\right)\right]\right\}, \quad (36)$$

$$\eta_{nl}^{AC}(\omega)=\frac{3\tilde{G}_{12}^{q}(\omega)}{5\tilde{G}_{12}^{d}(\omega)\left[(l+1)F_{l+1}^{2}\tilde{D}_{l+1}^{2}+lF_{l-1}^{2}\tilde{D}_{l-1}^{2}\right]}\times$$
$$\times\left\{\frac{(l+1)(l+2)}{(2l+1)(2l+3)}F_{l+2}\tilde{Q}_{l+2}\left[5lF_{l-1}\tilde{D}_{l-1}\tilde{D}_{l-1}\cos\left(\tilde{\tilde{\delta}}_{l+2}-\tilde{\tilde{\delta}}_{l-1}\right)-\right.\right.$$
$$\left.-(l+3)F_{l+1}\tilde{D}_{l+1}\cos\left(\tilde{\tilde{\delta}}_{l+2}-\tilde{\tilde{\delta}}_{l-1}\right)\right]-\frac{(l-1)l}{(2l+1)(2l+1)}F_{l-2}\tilde{Q}_{l-2}\times \quad (37)$$
$$\times\left[5(l+1)F_{l+1}\tilde{D}_{l+1}\cos\left(\tilde{\tilde{\delta}}_{l-2}-\tilde{\tilde{\delta}}_{l+1}\right)-(l-2)F_{l-1}\tilde{D}_{l-1}\cos\left(\tilde{\tilde{\delta}}_{l-2}-\tilde{\tilde{\delta}}_{l-1}\right)\right]+$$
$$\left.+2\frac{l(l+1)F_{l}\tilde{Q}_{l}}{(2l-1)(2l+3)}\left[(l+2)F_{l+1}\tilde{D}_{l+1}\tilde{D}_{l+1}\cos\left(\tilde{\tilde{\delta}}_{l}-\tilde{\tilde{\delta}}_{l+1}\right)-(l-1)F_{l-1}\tilde{D}_{l-1}\tilde{D}_{l-1}\cos\left(\tilde{\tilde{\delta}}_{l}-\tilde{\tilde{\delta}}_{l-1}\right)\right]\right\}.$$

They include also quadrupole photoionization matrix elements $Q_{l,l\pm2}$ that, being complex numbers can be presented as $Q_{l,l\pm2}(\omega)\equiv\tilde{Q}_{l,l\pm2}(\omega)\exp[i\Delta_{l,l\pm2}(k)]$, where $\tilde{Q}_{l,l\pm2}(\omega)$ is the module of $Q_{l,l\pm2}(\omega)$.

Very often experimentalists are using non-dipole parameters $\gamma_{nl}^{C}$ and $\delta_{nl}^{C}$, introduced in [22, 23]. The following formula connect them to those defined by (36) and (37)

$$\gamma_{nl}^{C}/5+\delta_{nl}^{C}=\kappa\gamma_{nl}, \qquad \gamma_{nl}^{C}/5=-\kappa\eta. \quad (38)$$

The dipole polarization amplitude factor is determined by (25), while the quadrupole one is given by similar to (25) relation

$$Q_{AC}(\omega)\cong Q_{A}(\omega)\left[1-\left(\frac{\alpha_{1}^{q}}{4R_{1}^{5}}+\frac{\alpha_{2}^{q}}{4R_{2}^{5}}\right)\frac{1-\dfrac{\alpha_{1}^{q}\alpha_{2}^{q}}{4\alpha_{1}^{q}R_{2}^{5}+4\alpha_{2}^{q}R_{1}^{5}}\left(1+\dfrac{R_{1}^{5}}{R_{2}^{5}}\right)}{1-\dfrac{\alpha_{1}^{q}\alpha_{2}^{q}}{16R_{2}^{6}}}\right]\equiv G_{12}^{q}(\omega)Q_{A}(\omega), \quad (39)$$

where $\alpha_{1}^{q}$ and $\alpha_{2}^{q}$ are the dynamic quadrupole polarizabilities of the fullerene inner and outer, respectively. Note, that that for a single-wall fullerene the quadrupole polarization amplitude factor is given by the relation [18]

$$Q_{AC}(\omega)\cong Q_{A}(\omega)\left(1-\frac{\alpha_{q}(\omega)}{4R^{5}}\right)\equiv Q_{A}(\omega)G^{q}(\omega), \quad (40)$$

Being complex numbers, the polarization amplitude factors $G_{12}^{q,d}(\omega)$ can be presented as

$$G_{12}^{q,d}(\omega) = \tilde{G}_{12}^{q,d}(\omega)\exp[i\Lambda^{q,d}(\omega)], \tag{41}$$

where $\tilde{G}_{12}^{q,d}(\omega)$ are the module of $G_{12}^{q,d}(\omega)$.

In (36) and (37) the following notations for phases are used $\tilde{\tilde{\delta}}_{l\pm1} = \tilde{\delta}_{l\pm1} + \Lambda^d = \delta_{l\pm1} + \Delta_{l\pm1} + \Lambda^d$ and $\tilde{\tilde{\delta}}_{l\pm2,l} = \tilde{\delta}_{l\pm2,l} + \Lambda^q = \delta_{l\pm2,l} + \Delta_{l\pm2,l} + \Lambda^q$.

If the reflection of photoelectrons by the fullerene shell is strong, it is insufficient to take it into account in the lowest order in $F_{l'}(k)$ as in (32). Natural to do it within the RPAE frame that is achieved by solving the following equation for the dipole amplitude

$$\left\langle \nu_1 \middle| \breve{D}(\omega) \middle| \nu_2 \right\rangle = \left\langle \nu_1 \middle| \hat{d} \middle| \nu_2 \right\rangle + \sum_{\nu_3,\nu_4} \frac{\left\langle \nu_3 \middle| \breve{D}(\omega) \middle| \nu_4 \right\rangle \left[ F_{\nu_3}^2 n_{\nu_4}(1-n_{\nu_3}) - F_{\nu_4}^2 n_{\nu_3}(1-n_{\nu_4}) \right] \left\langle \nu_4\nu_1 \middle| U \middle| \nu_3\nu_2 \right\rangle}{\varepsilon_{\nu_4} - \varepsilon_{\nu_3} + \omega + i\eta(1-2n_{\nu_3})} \tag{42}$$

and for the quadrupole amplitude

$$\left\langle \nu_1 \middle| \breve{Q}(\omega) \middle| \nu_2 \right\rangle = \left\langle \nu_1 | \hat{q} | \nu_2 \right\rangle + \sum_{\nu_3,\nu_4} \frac{\left\langle \nu_3 \middle| \breve{Q}(\omega) \middle| \nu_4 \right\rangle \left[ F_{\nu_3}^2 n_{\nu_4}(1-n_{\nu_3}) - F_{\nu_4}^2 n_{\nu_3}(1-n_{\nu_4}) \right] \left\langle \nu_4\nu_1 \middle| U \middle| \nu_3\nu_2 \right\rangle}{\varepsilon_{\nu_4} - \varepsilon_{\nu_3} + \omega + i\eta(1-2n_{\nu_3})}. \tag{43}$$

The relations (42, 43) are generalization of the RPAE equations for an atom surrounded by zero-thickness potential walls.

The relation (32) with pure atomic cross-section $\sigma_{nl,kl'}^{A}(\omega)$ is not valid any more. Instead, we have

$$\sigma_{nl,kl'}^{AC_{12}}(\omega) = |F_{l'}(k)|^2 \, S_{12}(\omega)\tilde{\sigma}_{nl,kl'}^{A}(\omega), \tag{44}$$

where $\tilde{\sigma}_{nl,kl'}^{A}(\omega)$ is given by relation $\tilde{\sigma}_{nl,kl'}^{A}(\omega) = \tilde{\sigma}_{nl,kl'}^{A,RPAE}(\omega)\,|\breve{D}_{nl,kl'}(\omega)|^2 / |D_{nl,kl'}(\omega)|^2$ with $D_{nl,kl'}(\omega)$ being the photoionization amplitude of an isolated atom in RPAE, while $\breve{D}_{nl,kl'}(\omega)$ is given by (42). Instead of (35) we have

$$\beta_{nl}(\omega) = \frac{1}{(2l+1)\left[ (l+1)F_{l+1}^2\tilde{\tilde{D}}_{l+1}^2 + lF_{l-1}^2\tilde{\tilde{D}}_{l-1}^2 \right]}[(l+1)(l+2)F_{l+1}^2\tilde{\tilde{D}}_{l+1}^2$$
$$+l(l-1)F_{l-1}^2\tilde{\tilde{D}}_{l-1}^2 - 6l(l+1)F_{l+1}F_{l-1}\tilde{\tilde{D}}_{l+1}\tilde{\tilde{D}}_{l-1}\cos(\tilde{\tilde{\delta}}_{l+1} - \tilde{\tilde{\delta}}_{l-1})] \quad . \tag{45}$$

By substituting RPAE amplitudes $D_{l\pm1}$ and $Q_{l,l\pm2}$ with $\breve{D}_{l\pm1}(\omega) \equiv \tilde{\tilde{D}}_{l\pm1}(\omega)\exp[i\breve{\Delta}_{l\pm1}(k)]$ and $\breve{Q}_{l,l\pm2}(\omega) \equiv \tilde{\tilde{Q}}_{l,l\pm2}(\omega)\exp[i\breve{\Delta}_{l,l\pm2}(k)]$ from (42) and (43) in (36-37) or (38), we obtain non-dipole angular anisotropy parameters with account of strong action of the fullerenes shell upon photoionization of two-shell endohedral atom. Note that corresponding phases $\tilde{\tilde{\delta}}_{l'}$ has to be substituted by $\tilde{\tilde{\tilde{\delta}}}_{l'}$ defined by the following relations $\tilde{\tilde{\tilde{\delta}}}_{l\pm1} = \tilde{\tilde{\delta}}_{l\pm1} + \Lambda^{d} = \delta_{l\pm1} + \breve{\Delta}_{l\pm1} + \Lambda^{d}$ and $\tilde{\tilde{\tilde{\delta}}}_{l\pm2,l} = \tilde{\tilde{\delta}}_{l\pm2,l} + \Lambda^{q} = \delta_{l\pm2,l} + \breve{\Delta}_{l\pm2,l} + \Lambda^{q}$ that take into account the difference between amplitudes $D_{l\pm1}$, $Q_{l,l\pm2}$ and $\breve{D}_{l\pm1}$, $\breve{Q}_{l,l\pm2}$, respectively.

To simplify the experimental detection of non-dipole parameters, the cross section (36) has to be measured under so-called *magic angle* equal to $\vartheta_m = 57.3^{\circ}$, for which $P_2(\cos\vartheta_m) = 0$, so that the contribution of the term д от члена с $\beta_{nl}$ отсутствует. Under angle $\vartheta_m$ the terms $\delta_{nl}^{C}(\omega)$ and $\gamma_{nl}^{C}(\omega)$ enter the cross section in the following combination:

$$\lambda_{nl}^{C} = \gamma_{nl}^{C} + 3\delta_{nl}^{C} \qquad (46)$$

There are no data available on the quadrupole polarizabilities. This is why we have decided to omit both the dipole and quadrupole polarization amplitude factors $G_{12}^{d}(\omega)$ and $G_{12}^{q}(\omega)$ in calculations of the non-dipole parameters.

## 7. Results of calculations

As concrete objects of two-shell endohedral atoms we consider here $Ar@C_{60}@C_{240}$ and $Xe@C_{60}@C_{240}$ with the fullerenes parameters, presented in Section 5. The following subshells are considered: $3p^6$ and $3s^2$ for Ar and $5p^6$, $5s^2$ and $4d^{10}$ for Xe. The results are presented in Fig. 1-23. They are obtained with the help of (42)-(45) and similarly modified (36-38, 46). The results with account of reflection by a single fullerenes shell we denote on the figures as FRPAE, while results with two shells taken into account are marked as FRPAE2. Specially mentioned is the effect of polarization factor $\bar{G}_{12}(\omega)$.

Since the reflection amplitudes $F_l(k)$ are different for different atoms, we are not presenting their values separately.

Fig. 1 and 2 depicts the $\bar{G}_{12}(\omega)$ parameter from (31) – its absolute value, real and imaginary parts. Fig. 2 shows, the G-factors separately for one shell or another. The G factor for the big fullerene is small but its combination with the inner shell enhances the common G-factor impressively. As it should be, effects of polarization are rapidly decreasing with $\omega$ growth, so that it approaches almost 1 at $\omega > 60\,\text{eV}$.

Fig.3 presents the photoionization cross-section for 3p in Ar, $Ar@C_{60}$, $Ar@C_{240}$ and $Ar@C_{60}@C_{240}$, i.e. with account of photoelectron scattering by one (a) and b)) and two (c) fullerenes shells. Fig. 4 demonstrates profound action of the polarization factor $G_{12}$ upon the Ar photoionization cross-section. Note that two-shell reflection concentrates almost all cross section into a single maximum. This maximum is strongly enhanced by fullerenes shells polarization.

Fig. 5 presents the same data as Fig.3 but for photoionization of 3s Ar. The action of polarization upon 3s is, as presented in Fig. 6, prominent but mach smaller than upon 3p. The effect of two-shell reflection even decreases the near threshold maximum.

Fig. 7 demonstrates the angular anisotropy parameter $\beta_{3p}(\omega)$ of 3p electrons in Ar, $Ar@C_{60}$ and $Ar@C_{60}@C_{240}$ with account of reflection factors F. The effect of reflection is quite small, leading to small oscillations around the photon energy region $\omega \approx 20\,\text{eV}$.

Fig. 8 depicts photoionization cross-section for 5p in Xe, $Xe@C_{60}$, $Xe@C_{60}@C_{240}$ and demonstrates profound action of the polarization factor $G_{12}$ upon the Xe 5p photoionization cross-section. The effect of scattering by the second fullerenes shell is strong enough, presenting a distinctive second maximum and decreasing the maximum that appear due to one-shell scattering. Polarization of fullerenes increases quite noticeable the cross section.

Fig. 9 presents the same data as Fig.8 but for photoionization of 5s in Xe. The role of polarization is strong at $\omega < 35\,\text{eV}$.

Fig. 10 depicts photoionization cross-section for 4d in Xe, Xe@$C_{60}$, Xe@$C_{60}$@$C_{240}$ and demonstrates action of the polarization factor $G_{12}$ upon the Xe 4d photoionization cross-section. Note that the role of polarization is very small and inclusion of two shells even suppresses the effect of only $C_{60}$.

Fig.11 demonstrates the dipole angular anisotropy parameter $\beta_{5p}(\omega)$ of 5p electrons in Xe, Xe@$C_{60}$ and Xe@$C_{60}$@$C_{240}$ with account of reflection factors F. The effect is similar in size to that in Fig. 7, but there are two regions of oscillations in 5p Xe instead of one in 3p Ar.

Fig.12 presents the dipole angular anisotropy parameter $\beta_{4d}(\omega)$ of 4d electrons in Xe, Xe@$C_{60}$ and Xe@$C_{60}$@$C_{240}$ with account of reflection factors F. Its effect is stronger than in Fig. 7 and 11 and located in an area, where the relative effect is bigger.

Fig. 13 gives the non-dipole angular anisotropy parameter $\gamma_{3p}^{C}(\omega)$ of 3p electrons in Ar, Ar@$C_{60}$ and Ar@$C_{60}$@$C_{240}$ with account of reflection factors F. It is remarkable that while the $C_{60}$ shell adds only relatively small oscillations to $\gamma_{3p}^{C}(\omega)$ as compared to the value in an isolated atom, the additional $C_{240}$ shell leads to an almost complete "mirror reflection" of the respective free atom curve with noticeable oscillations on it.

Fig. 14 demonstrates the non-dipole angular anisotropy parameter $\delta_{3p}^{C}(\omega)$ of 3p electrons in Ar, Ar@$C_{60}$ and Ar@$C_{60}$@$C_{240}$ with account of reflection factors F. Qualitatively, the situation is similar to that in Fig. 13, but the absolute values are by a factor of three smaller.

Fig.15 depicts the combination of non-dipole angular anisotropy parameters $\gamma_{3p}^{C}(\omega)+3\delta_{3p}^{C}(\omega)$ of 3p electrons in Ar, Ar@$C_{60}$ and Ar@$C_{60}$@$C_{240}$ with account of reflection factors F. The inclusion of the second fullerenes shell apart of "mirror reflection" adds prominent structure to the curve.

Fig. 16 gives the non-dipole angular anisotropy parameter $\gamma_{3s}^{C}(\omega)$ of 3s electrons in Ar, Ar@$C_{60}$ and Ar@$C_{60}$@$C_{240}$ with account of reflection factors F. The role of fullerenes shells is much weaker than for 3p electrons.

Fig. 17 depicts the non-dipole angular anisotropy parameter $\gamma_{5p}^{C}(\omega)$ of 5p electrons in Xe, Xe@$C_{60}$ and Xe@$C_{60}$@$C_{240}$ with account of reflection factors F. While inclusion of the $C_{60}$ shell to some extend enhances the parameter, the second shell brings this parameter almost to zero except a noticeable maximum at $\omega \approx 20$ eV .

Fig. 18 shows the non-dipole angular anisotropy parameter $\delta_{5p}^{C}(\omega)$ of 5p electrons in Xe, Xe@$C_{60}$ and Xe@$C_{60}$@$C_{240}$ with account of reflection factors F. The situation is similar to that for $\gamma_{5p}^{C}(\omega)$.

Fig. 19 presents the combination of non-dipole angular anisotropy parameters $\gamma_{3p}^{C}(\omega)+3\delta_{3p}^{C}(\omega)$ for 5p electrons in Xe, Xe@$C_{60}$ and Xe@$C_{60}$@$C_{240}$ with account of reflection factors F. The situation is similar to that in Fig. 17 and 18, since summation of zeroes leads to a zero. An outstanding structure is a maximum at the same place as for $\gamma_{5p}^{C}(\omega)$.

Fig. 20 gives the non-dipole angular anisotropy parameter $\gamma_{5s}^{C}(\omega)$ of 5s electrons in Xe, Xe@$C_{60}$ and Xe@$C_{60}$@$C_{240}$ with account of reflection factors F. On the low $\omega$ side inclusion of fullerenes shells affects the parameter considerably but at $\omega \geq 50\,\text{eV}$ this influence disappears.

Fig. 21 demonstrates the non-dipole angular anisotropy parameter $\gamma_{4d}^{C}(\omega)$ of 4d electrons in Xe, Xe@$C_{60}$ and Xe@$C_{60}$@$C_{240}$ with account of reflection factors F. The $C_{60}$ shell produces an oscillation at 150-175 eV, while the second fullerene shell makes the curve where it is non-zero an almost complete "mirror reflection" of the free atom curve.

Fig. 22 gives the non-dipole angular anisotropy parameter $\gamma_{4d}^{C}(\omega)$ of 4d electrons in Xe, Xe@$C_{60}$ and Xe@$C_{60}$@$C_{240}$ with account of reflection factors F. Already $C_{60}$ changes the sign of the parameter and ads an oscillation. The addition of $C_{240}$ brings again to a curve similar to "mirror reflection" of the isolated atom curve.

Fig. 23 depicts the combination of non-dipole angular anisotropy parameters $\gamma_{4d}^{C}(\omega)+3\delta_{4d}^{C}(\omega)$ of 4d electrons in Xe, Xe@$C_{60}$ and Xe@$C_{60}$@$C_{240}$ with account of reflection factors F. The modification due to two shells is most visible at $\omega > 140\,\text{eV}$, where the magnitude becomes smaller and at $\omega > 180\,\text{eV}$ changes sign.

## 8. Conclusion

We present results for photoionization cross sections, dipole and non-dipole angular anisotropy parameters for outer in Ar and outer and intermediate in Xe subshells for double caged atoms Ar@$C_{60}$@$C_{240}$ and Xe@$C_{60}$@$C_{240}$ and compared it to the respective data for isolated atoms. We have investigated effects of photoelectron scattering by two zero-thickness potential wells and modification of the incoming photon beam due to dipole polarization of both fullerenes shell.

A whole variety of resonances is found that obviously are far from being a simple sum of effects, given by single fullerenes shells. Particularly sensitive to the surrounding of fullerenes shell are non-dipole angular anisotropy parameters. For them, however, the effect of deviation of real fullerenes shell potential from the ideal spherical shape (1) could be of importance.

We admit that the moment when investigation of photoionization of such objects as A@$C_{60}$@$C_{240}$ or similar will take place is not literally tomorrow. However, we hope to see them being performed in not too distant future. We believe that possible findings will justify the efforts.

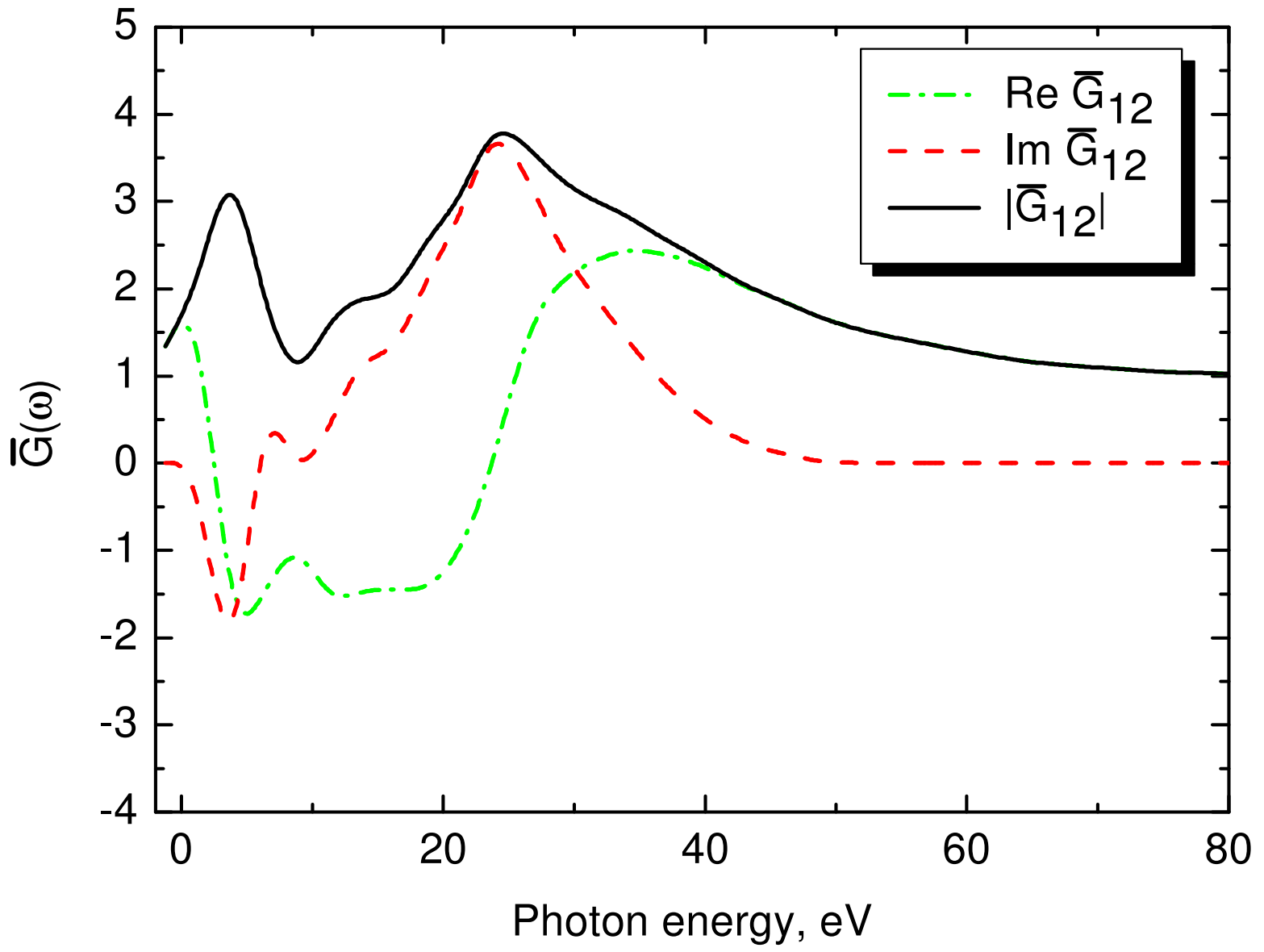

Fig.1. Polarization factor $\bar{G}_{12}(\omega)$ for $C_{60}$@$C_{240}$ – real, imaginary and absolute values.

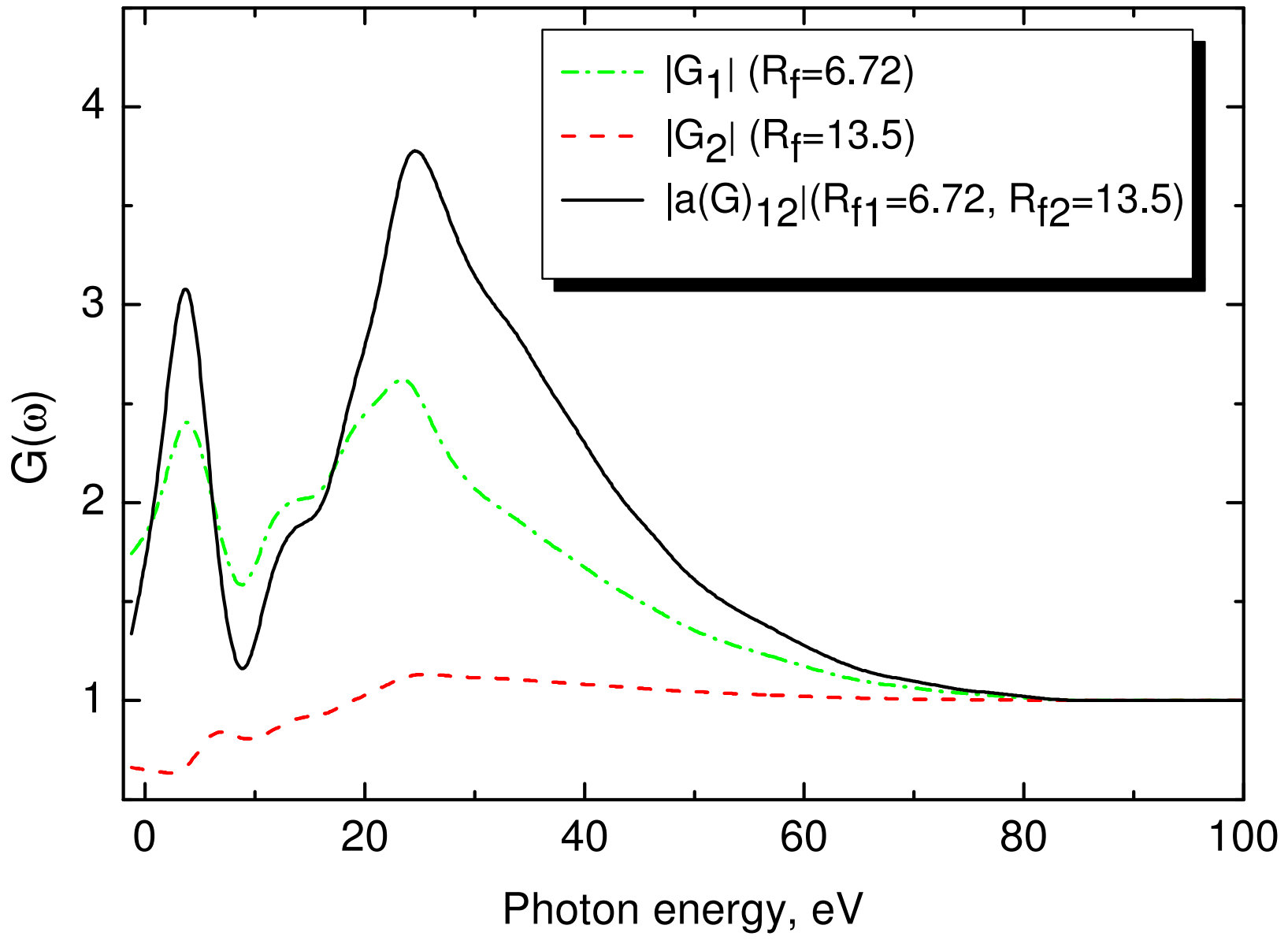

Fig.2. Polarization factors $|G_1(\omega)|$ for $C_{60}$, $|G_2(\omega)|$ for $C_{240}$ and $|\bar{G}_{12}(\omega)|$ for $C_{60}$@$C_{240}$.

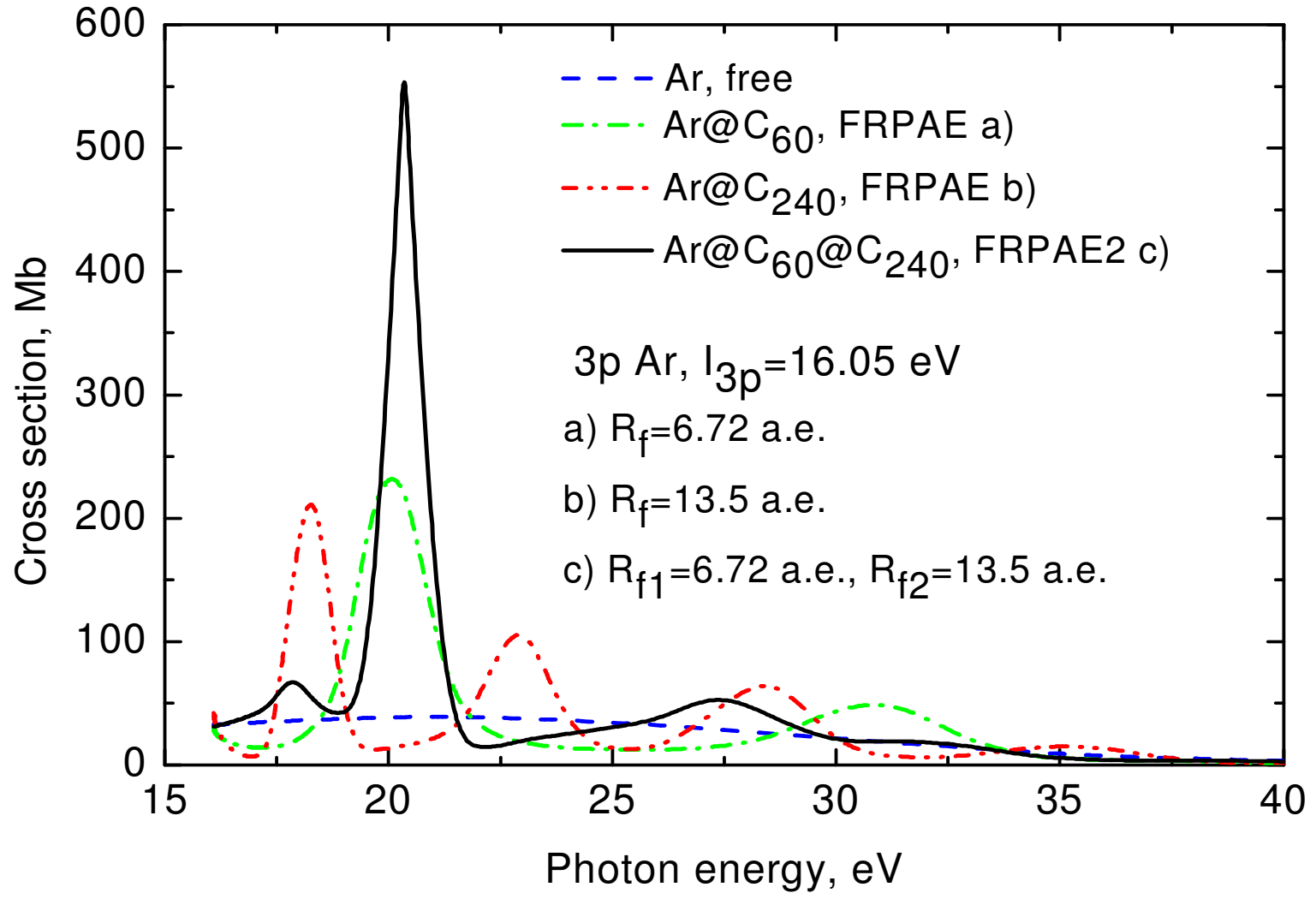

Fig3. Photoionization cross-section of 3p electrons in Ar, Ar@$C_{60}$ and Ar@$C_{60}$@$C_{240}$ with account of reflection factors F

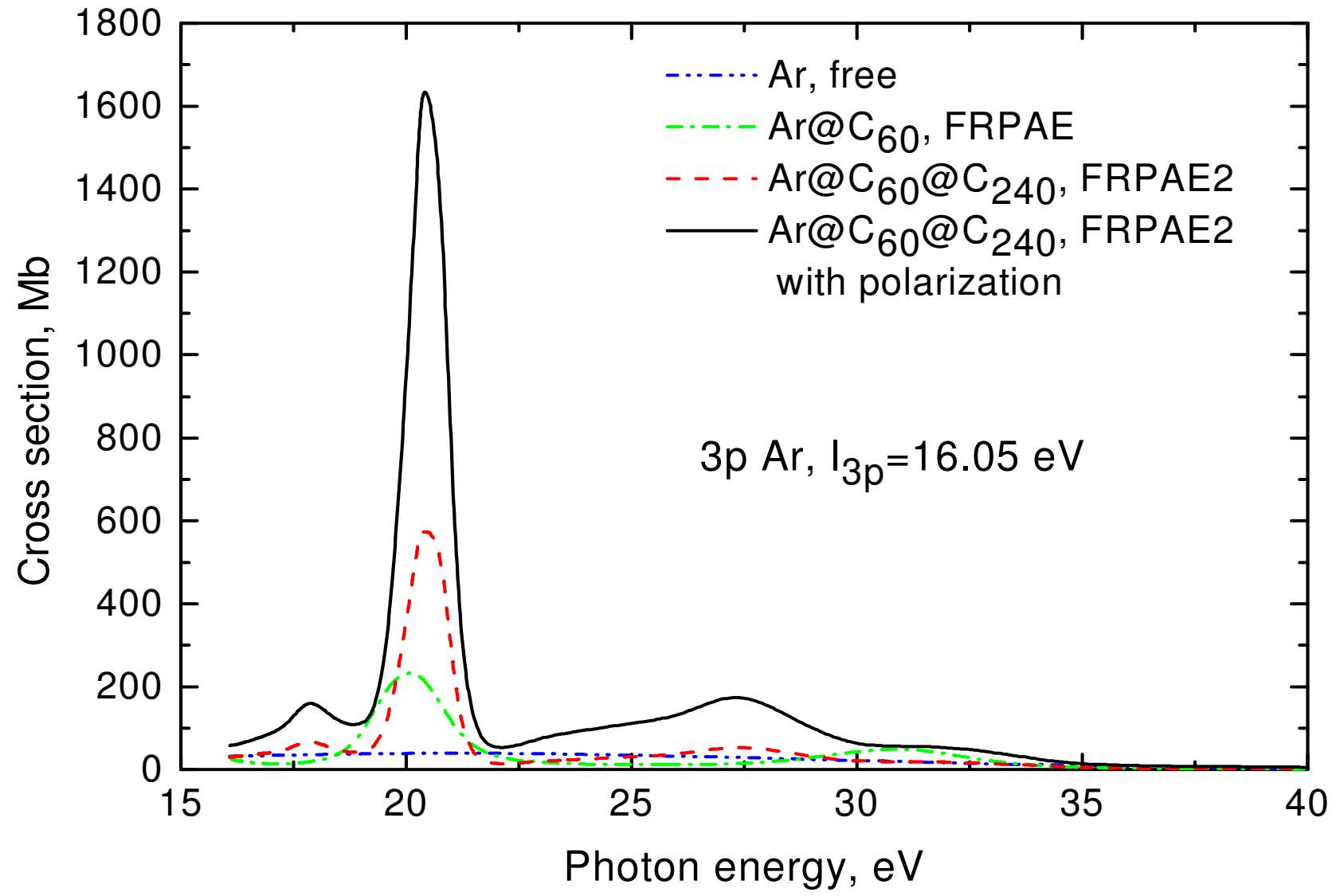

Fig.4. Photoionization cross-section of 3p electrons in Ar, Ar@$C_{60}$ and Ar@$C_{60}$@$C_{240}$ with account of reflection factors F and the latter with account of polarization factor $G_{12}(\omega)$.

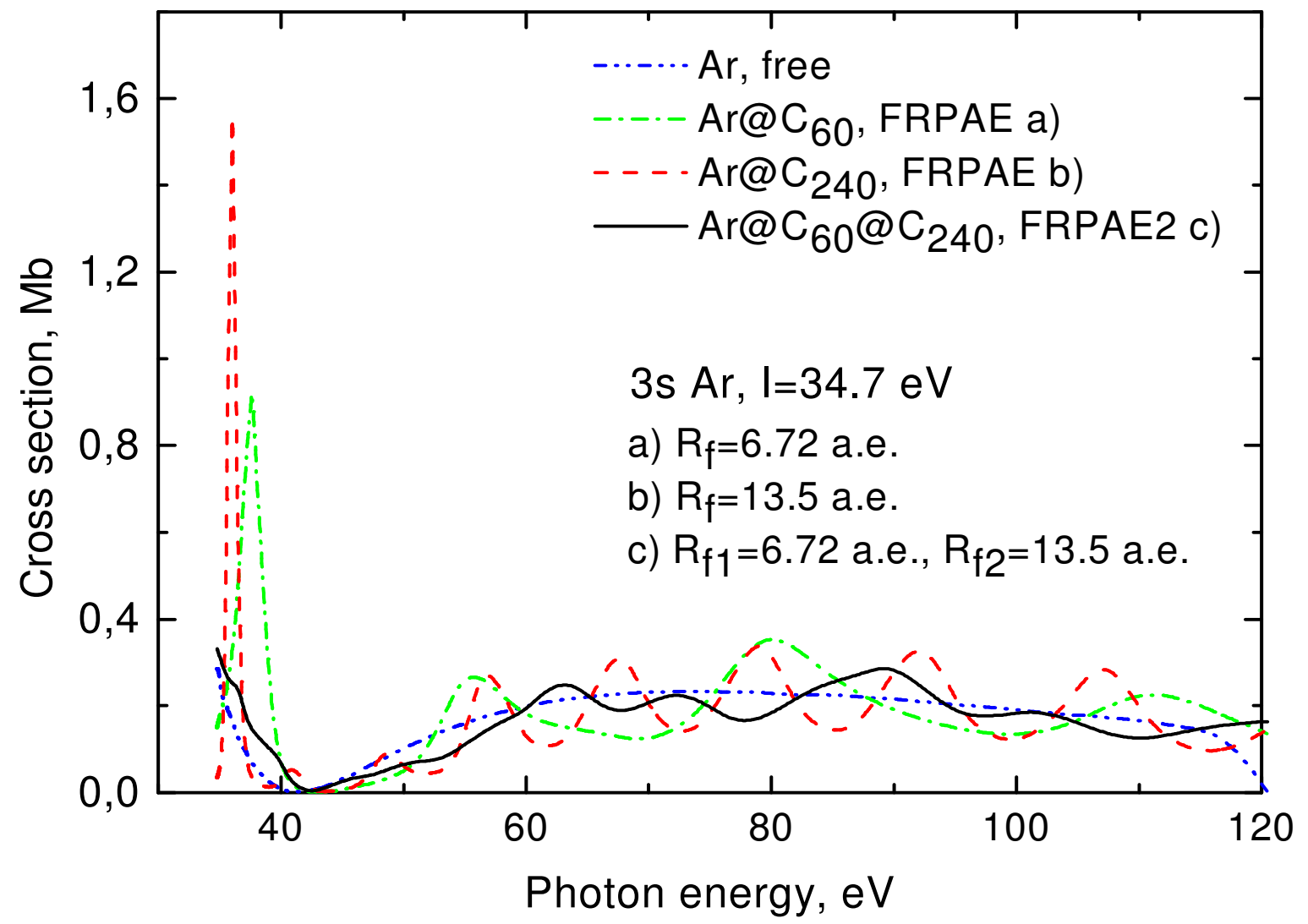

Fig. 5. Photoionization cross-section of 3s electrons in Ar, $Ar@C_{60}$, $Ar@C_{240}$ and $Ar@C_{60}@C_{240}$ with account of reflection factors F.

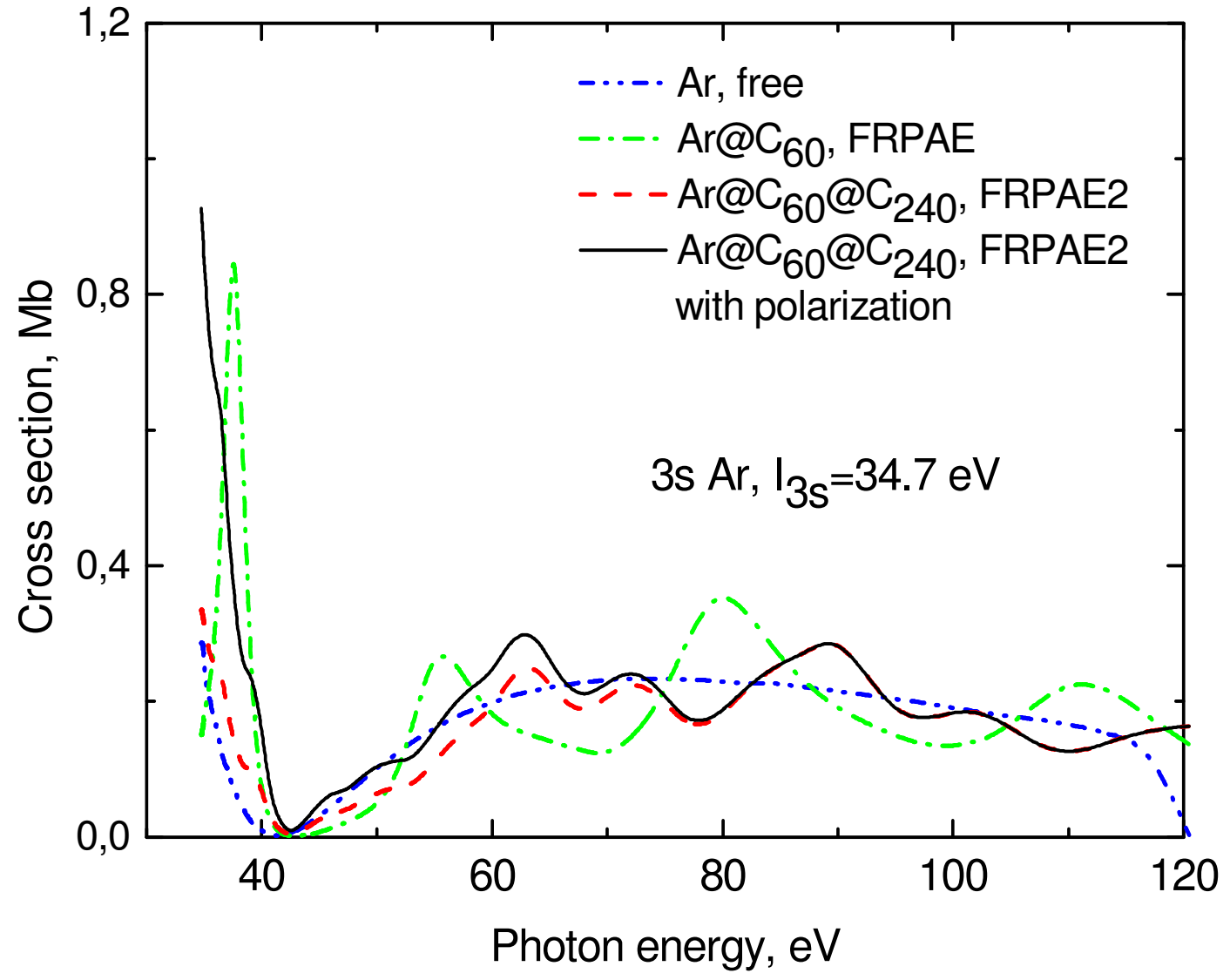

Fig.6. Photoionization cross-section of 3s electrons in Ar, $Ar@C_{60}$ and $Ar@C_{60}@C_{240}$ with account of reflection factors F and the latter with account of polarization factor $G_{12}(\omega)$.

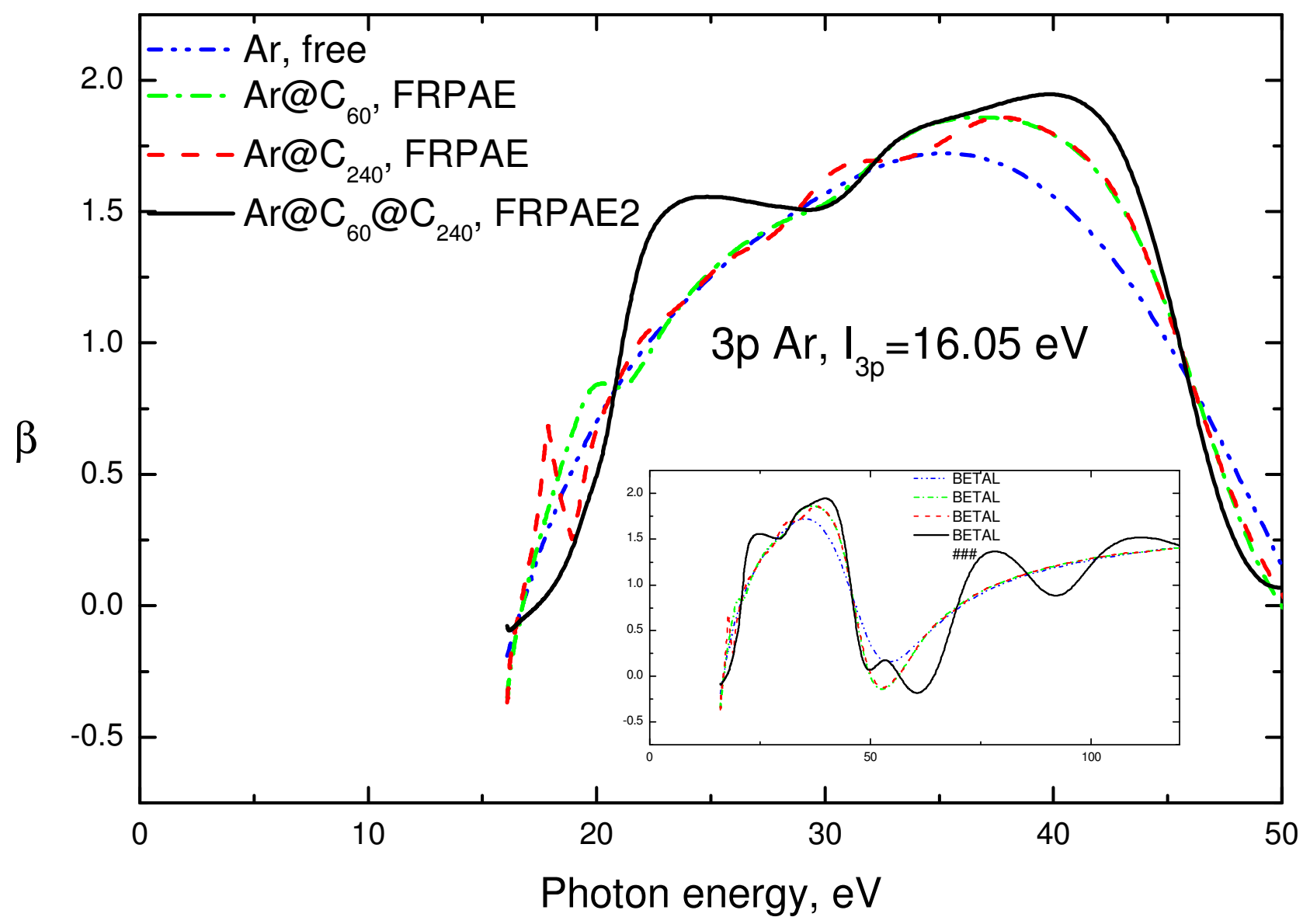

Fig. 7. Dipole angular anisotropy parameter $\beta_{3p}(\omega)$ of 3p electrons in Ar, $Ar@C_{60}$ and $Ar@C_{60}@C_{240}$ with account of reflection factors F.

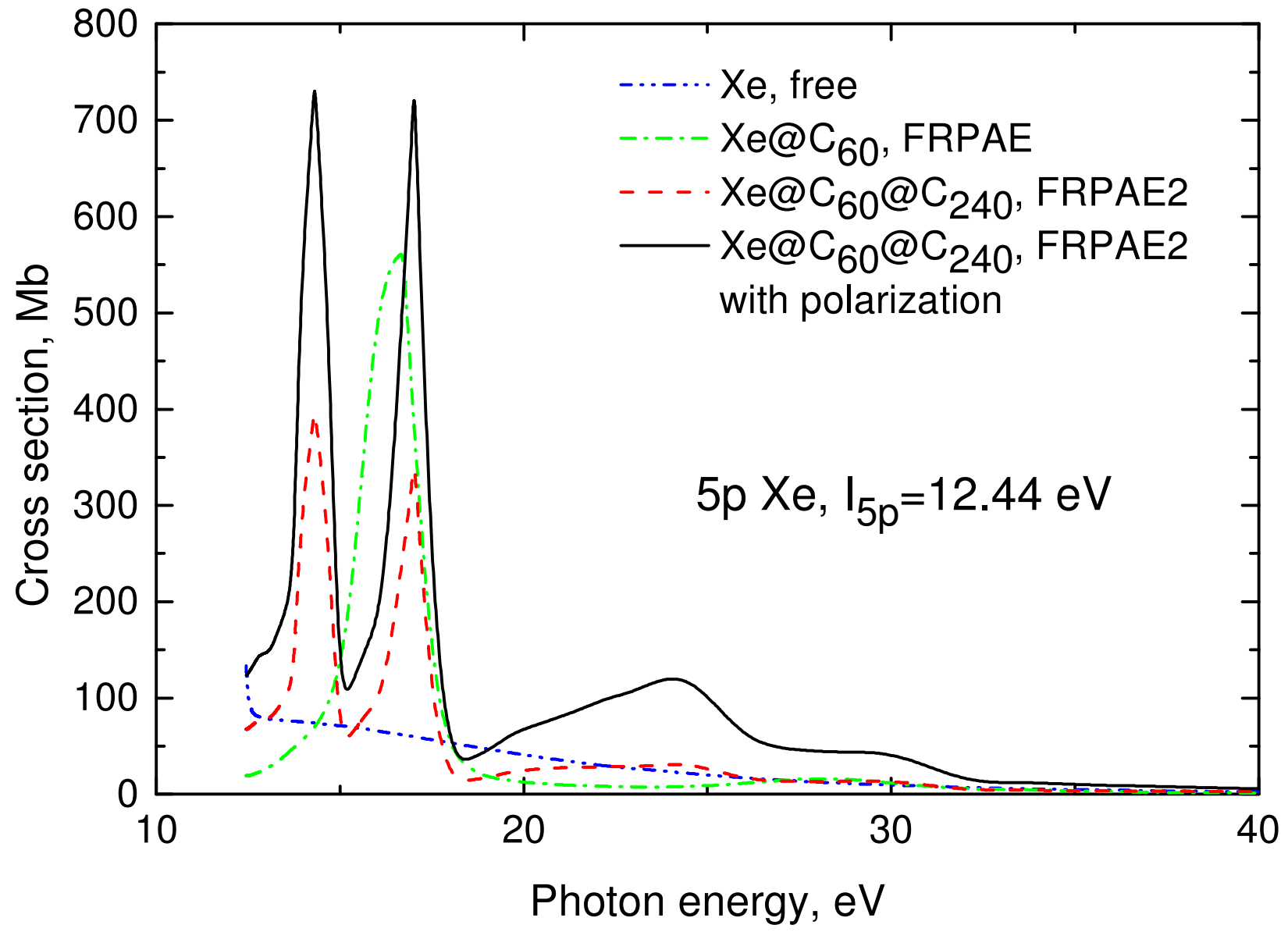

Fig.8. Photoionization cross-section of 5p electrons in Xe, $Xe@C_{60}$ and $Xe@C_{60}@C_{240}$ with account of reflection factors F and the latter with account of polarization factor $G_{12}(\omega)$.

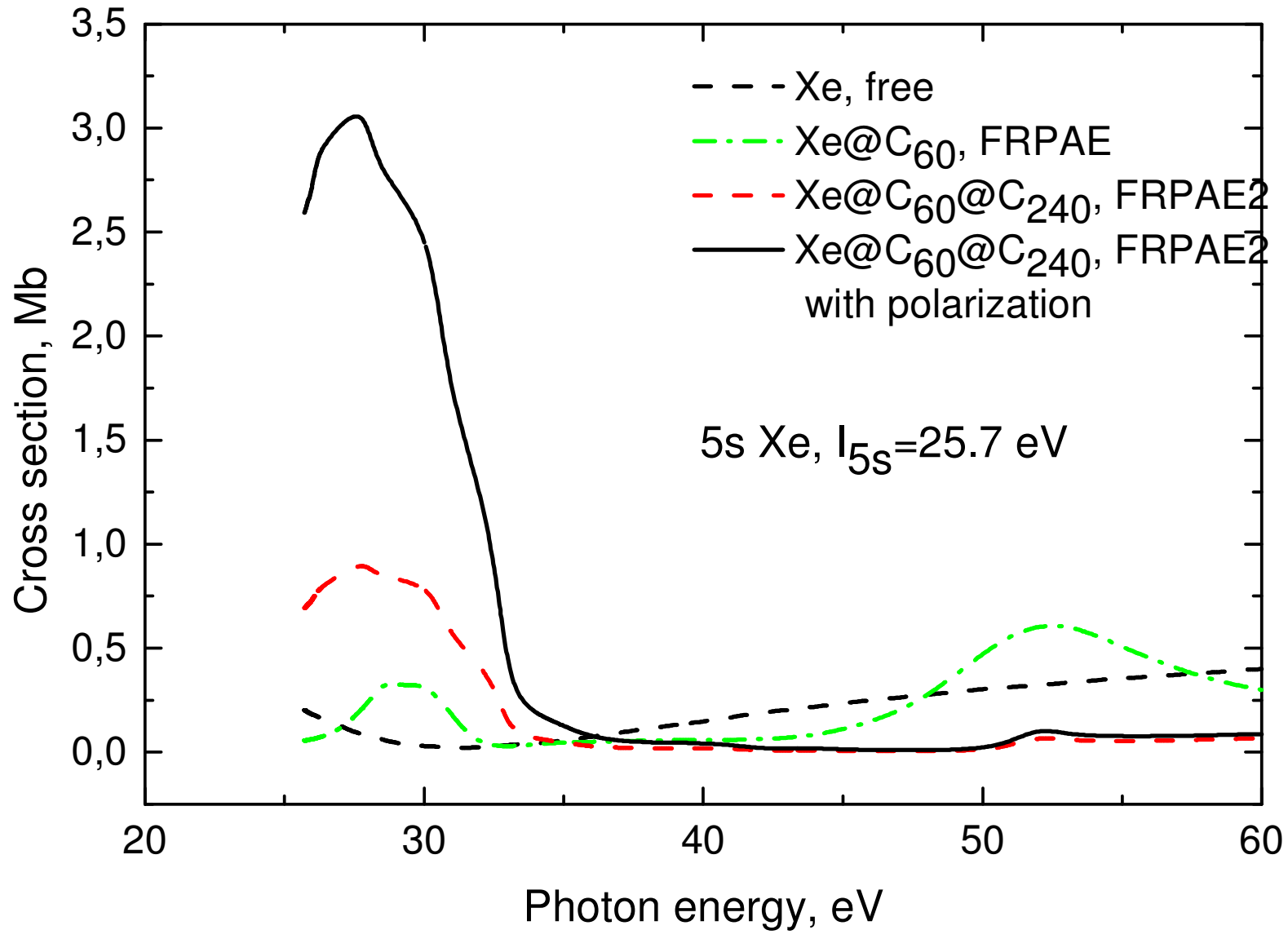

Fig.9. Photoionization cross-section of 5p electrons in Xe, Xe@$C_{60}$ and Xe@$C_{60}$@$C_{240}$ with account of reflection factors F and the latter with account of polarization factor $G_{12}(\omega)$.

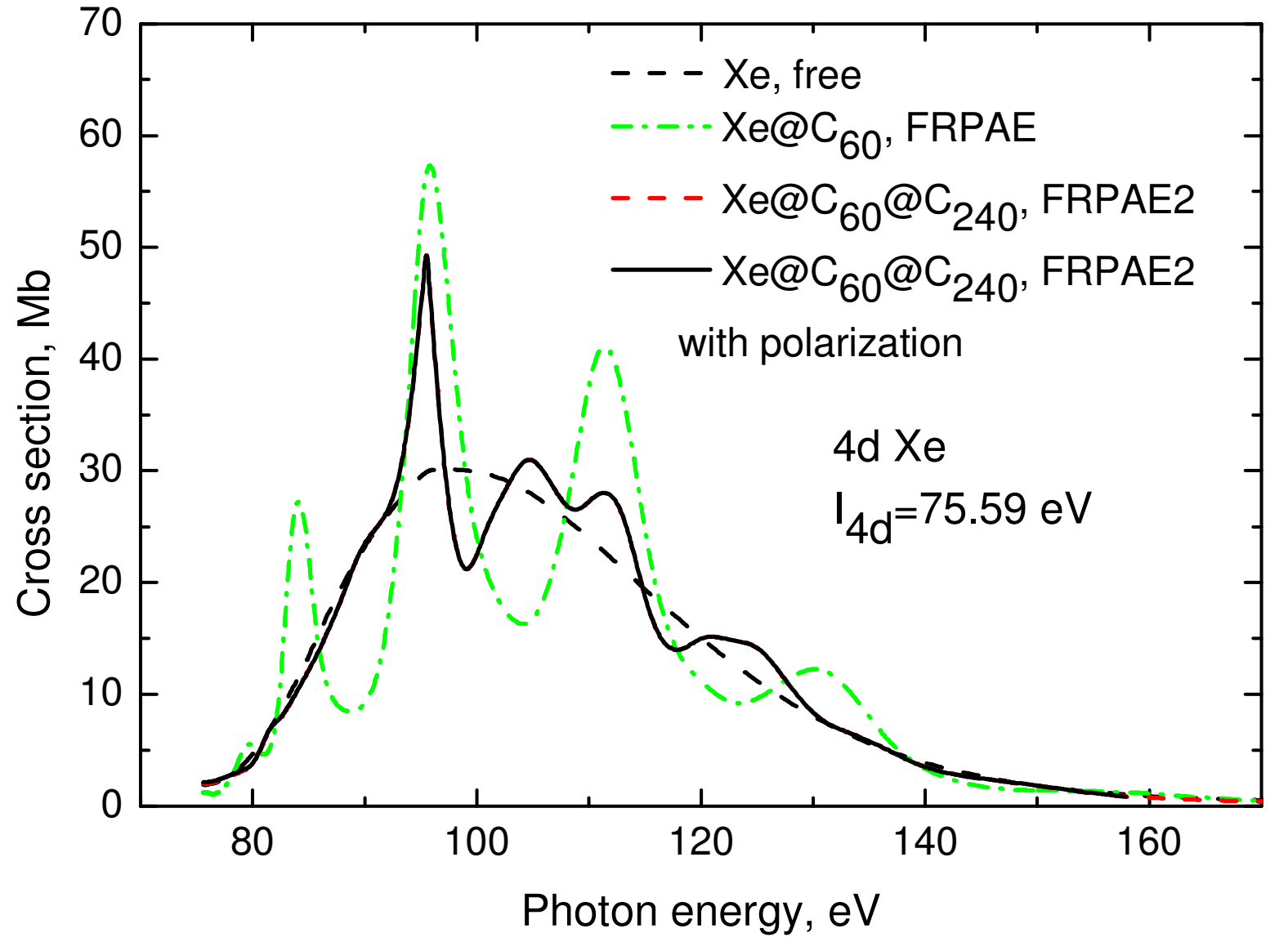

Fig.10. Photoionization cross-section of 4d electrons in Xe, Xe@$C_{60}$ and Xe@$C_{60}$@$C_{240}$ with account of reflection factors F and the latter with account of polarization factor $G_{12}(\omega)$.

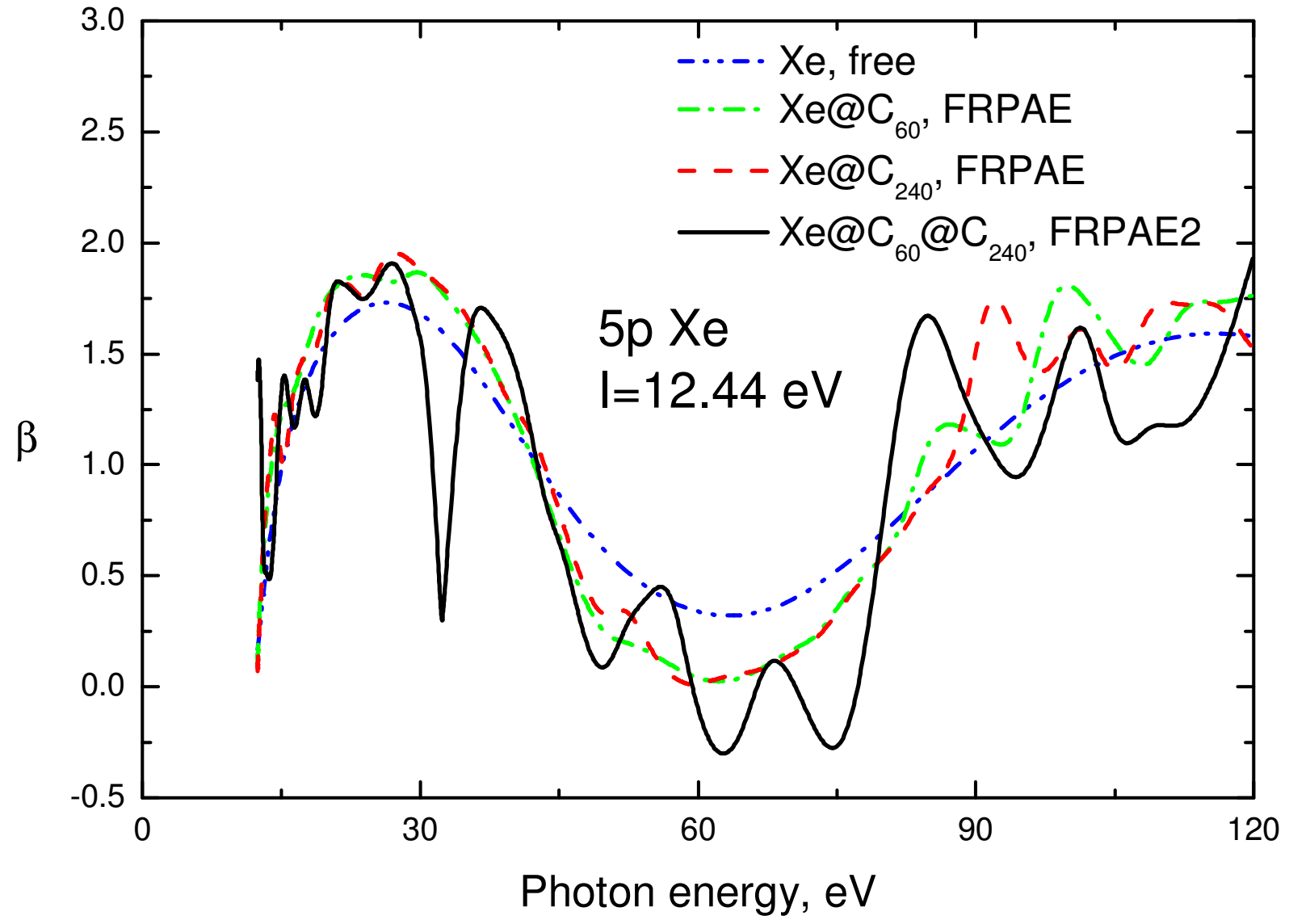

Fig.11. Dipole angular anisotropy parameter $\beta_{5p}(\omega)$ of 5p electrons in Xe, Xe@$C_{60}$ and Xe@$C_{60}$@$C_{240}$ with account of reflection factors F.

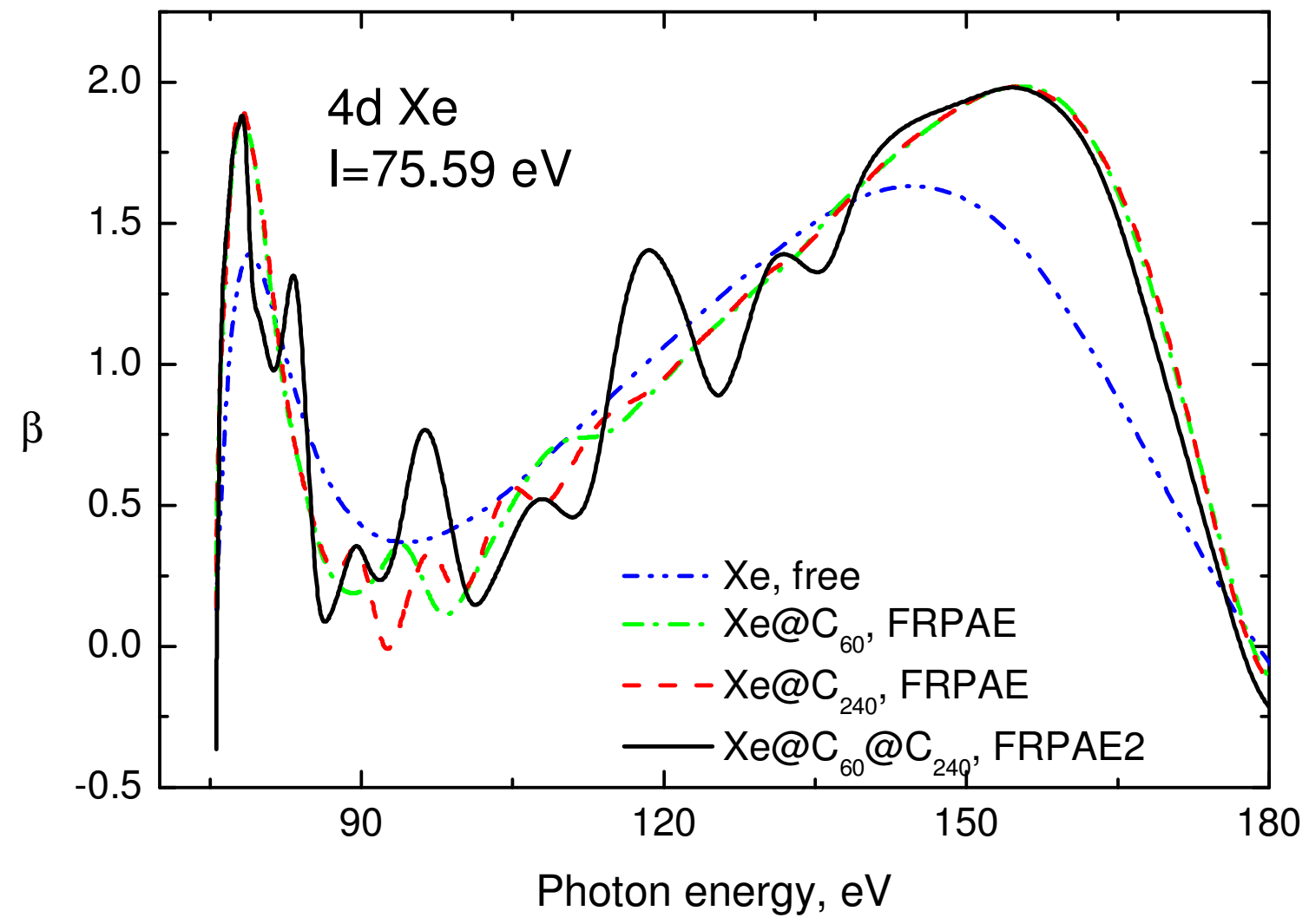

Fig.12. Dipole angular anisotropy parameter $\beta_{4d}(\omega)$ of 4d electrons in Xe, Xe@$C_{60}$ and Xe@$C_{60}$@$C_{240}$ with account of reflection factors F.

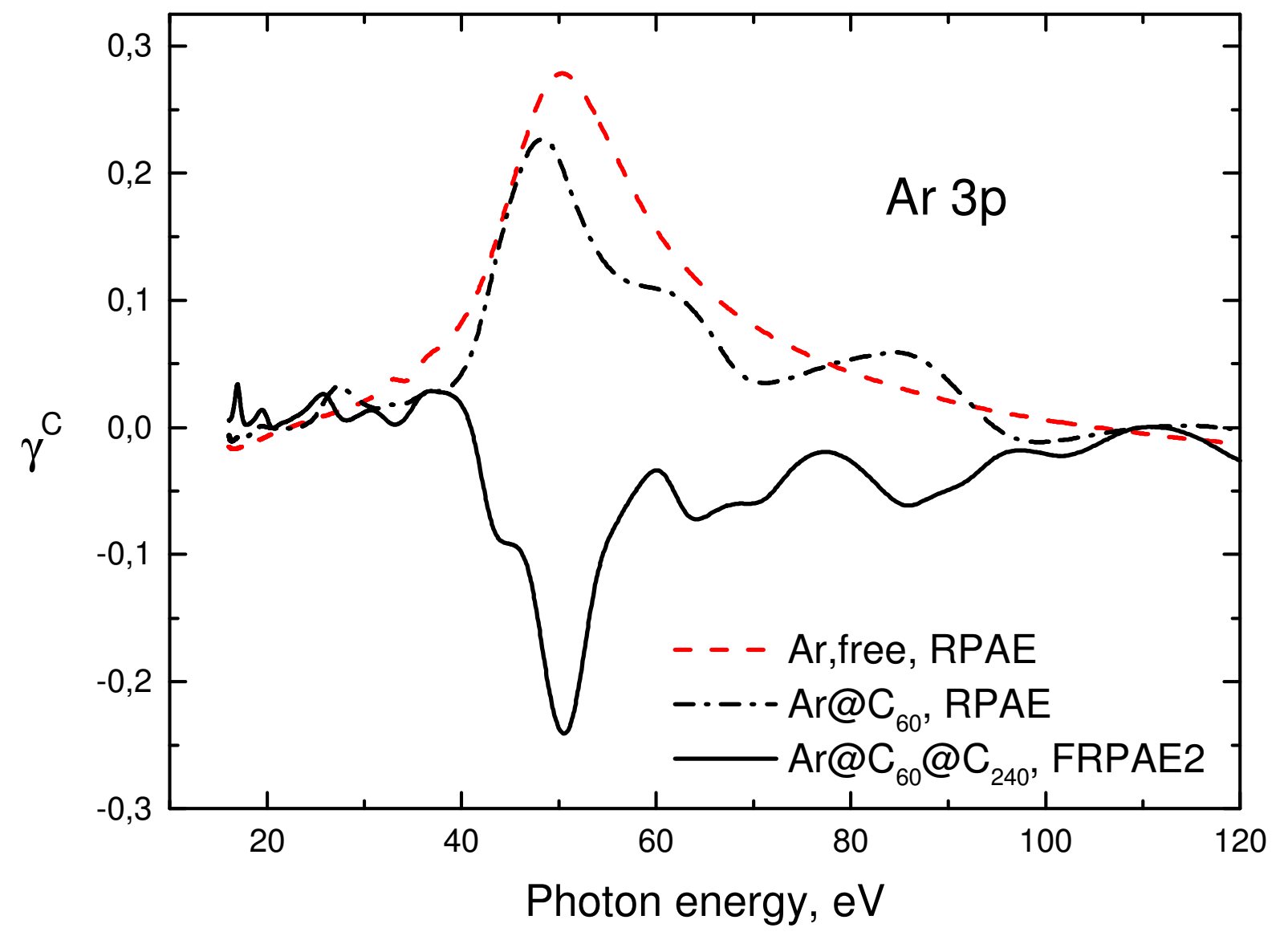

Fig.13. Non-dipole angular anisotropy parameter $\gamma^{C}_{3p}(\omega)$ of 3p electrons in Ar, Ar@$C_{60}$ and Ar@$C_{60}$@$C_{240}$ with account of reflection factors F.

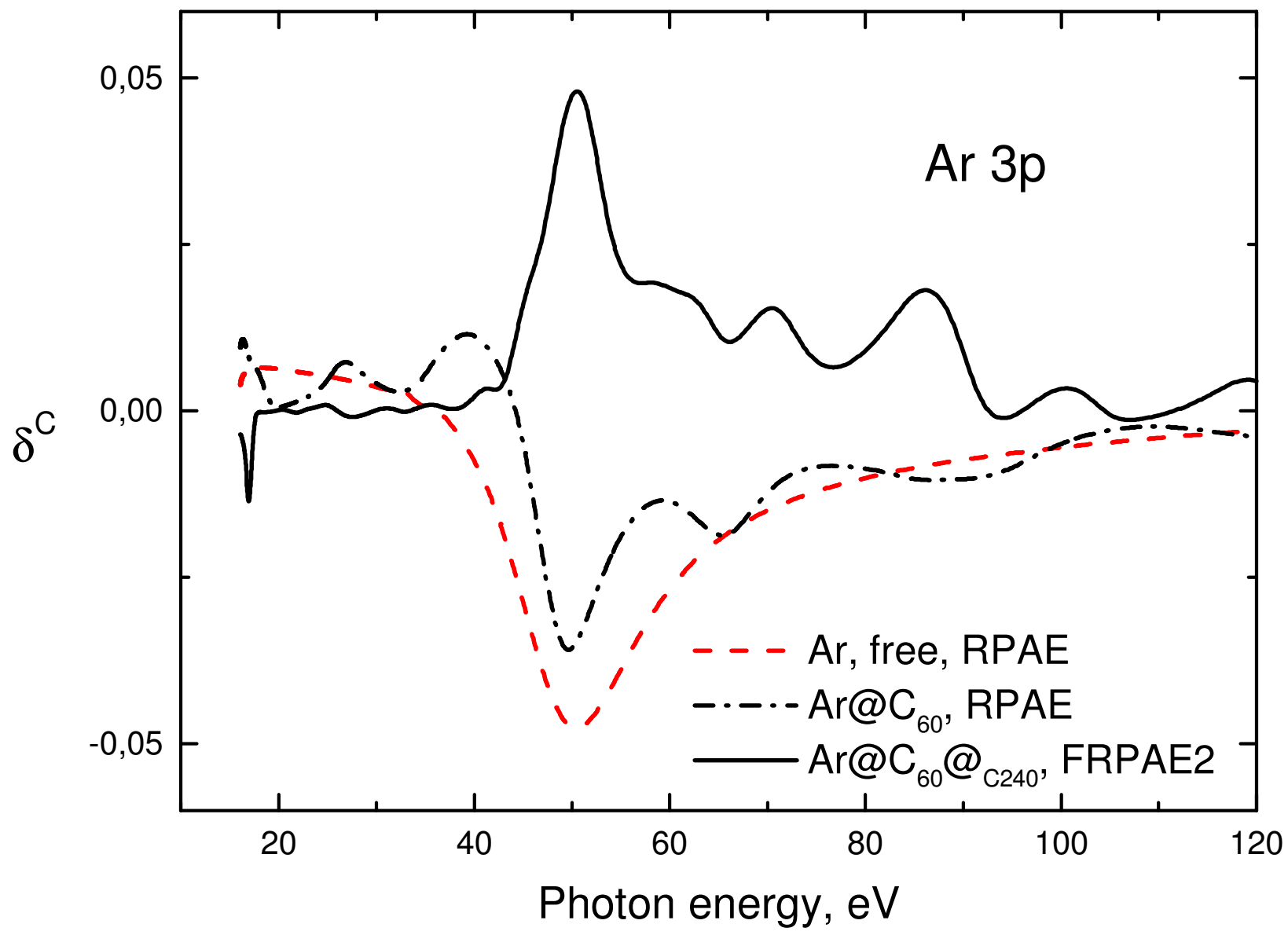

Fig.14. Non-dipole angular anisotropy parameter $\delta^{C}_{3p}(\omega)$ of 3p electrons in Ar, Ar@$C_{60}$ and Ar@$C_{60}$@$C_{240}$ with account of reflection factors F.

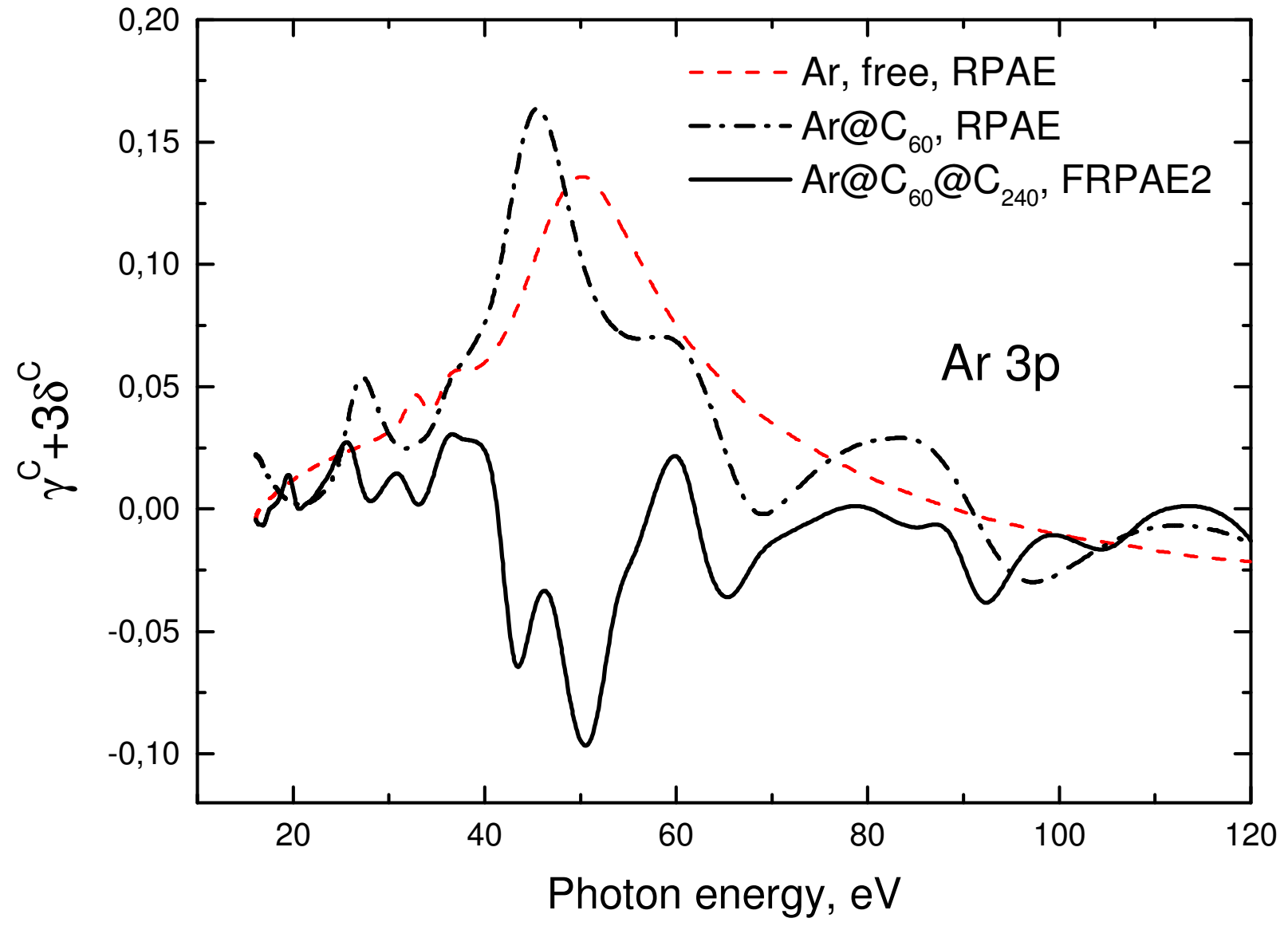

Fig.15. Combination of non-dipole angular anisotropy parameters $\gamma_{3p}^{C}(\omega)+3\delta_{3p}^{C}(\omega)$ of 3p electrons in Ar, Ar@$C_{60}$ and Ar@$C_{60}$@$C_{240}$ with account of reflection factors F.

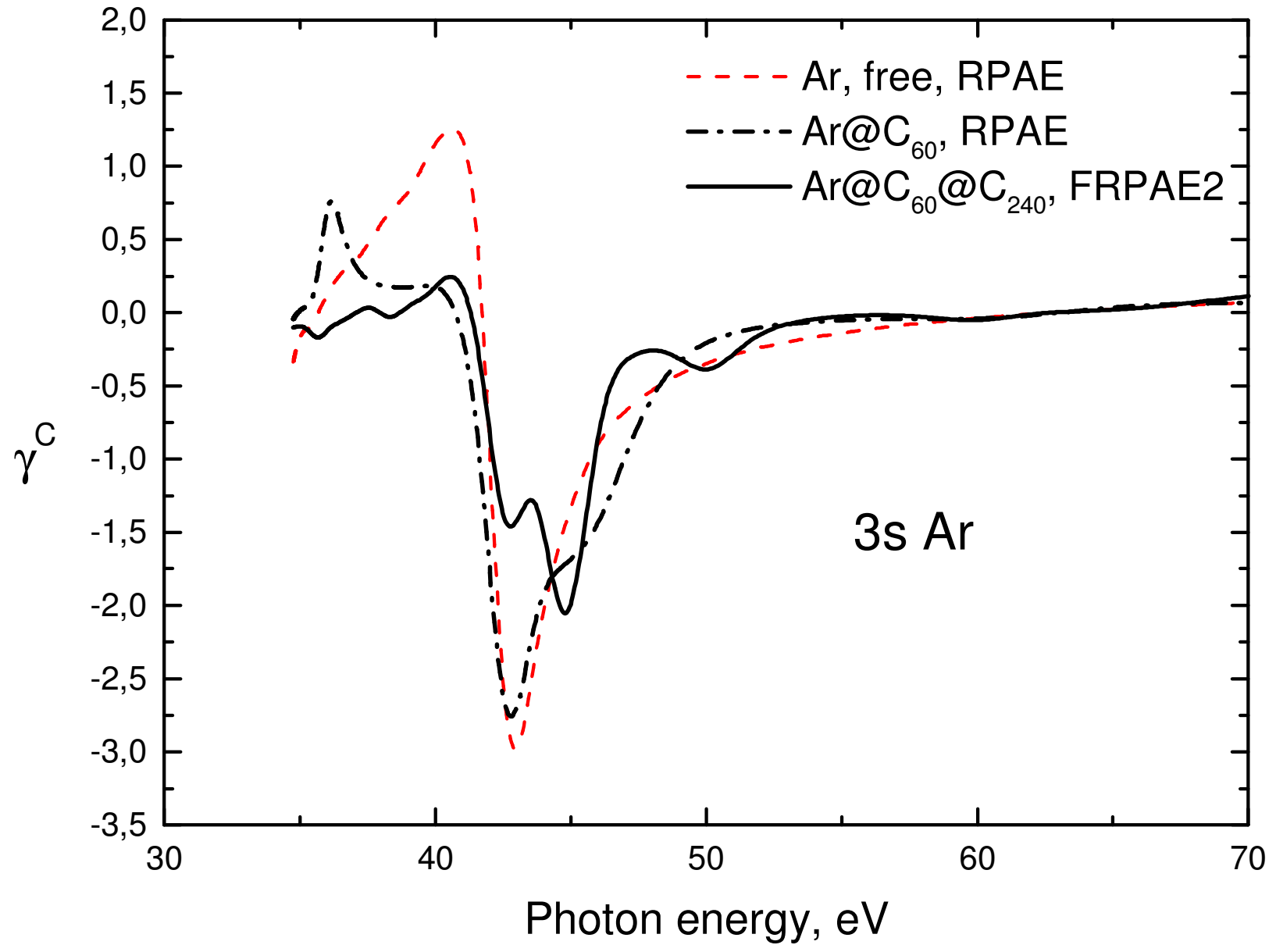

Fig.16. Non-dipole angular anisotropy parameter $\gamma_{3s}^{C}(\omega)$ of 3s electrons in Ar, Ar@$C_{60}$ and Ar@$C_{60}$@$C_{240}$ with account of reflection factors F.

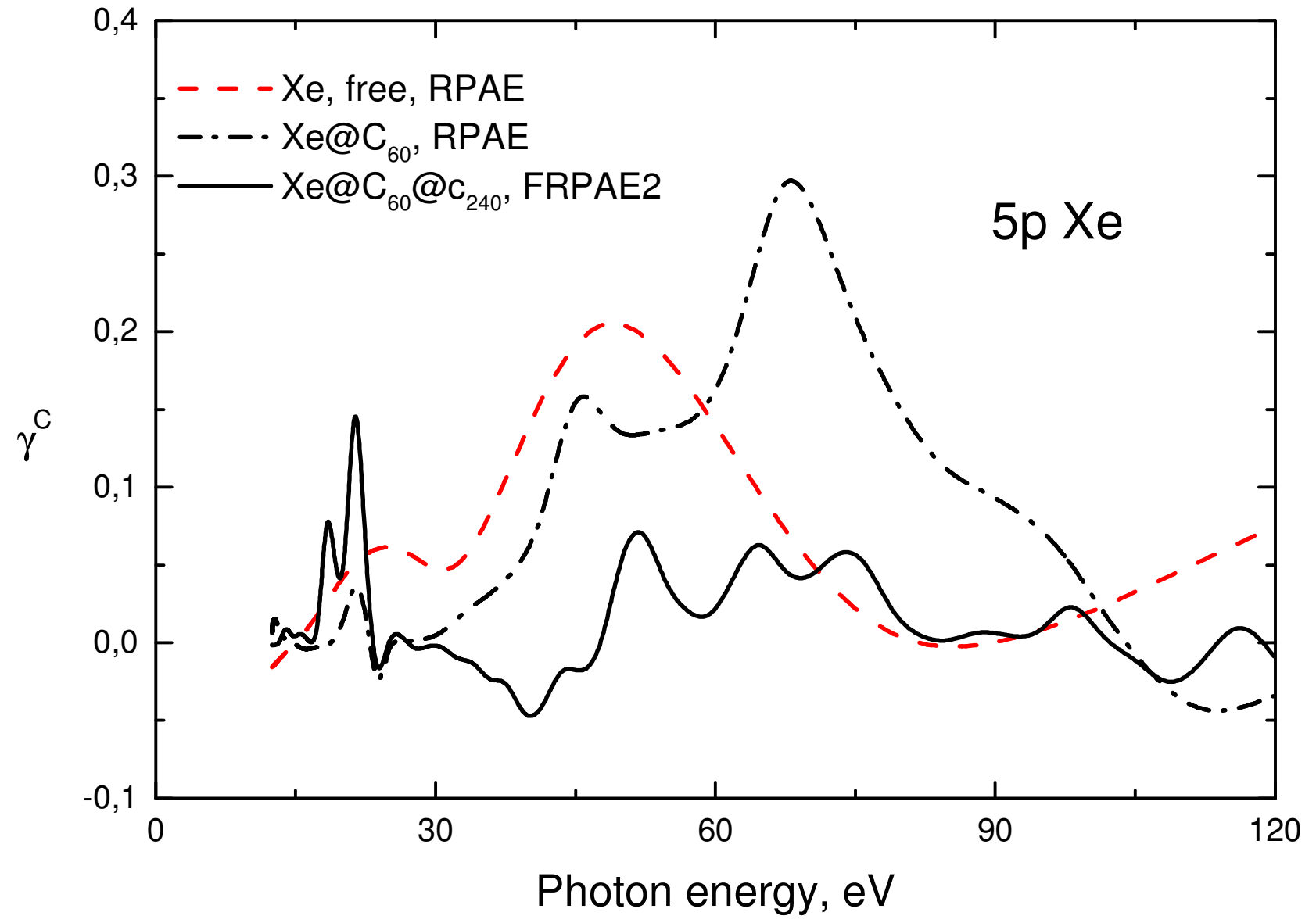

Fig.17. Non-dipole angular anisotropy parameter $\gamma_{5p}^{C}(\omega)$ of 5p electrons in Xe, $Xe@C_{60}$ and $Xe@C_{60}@C_{240}$ with account of reflection factors F.

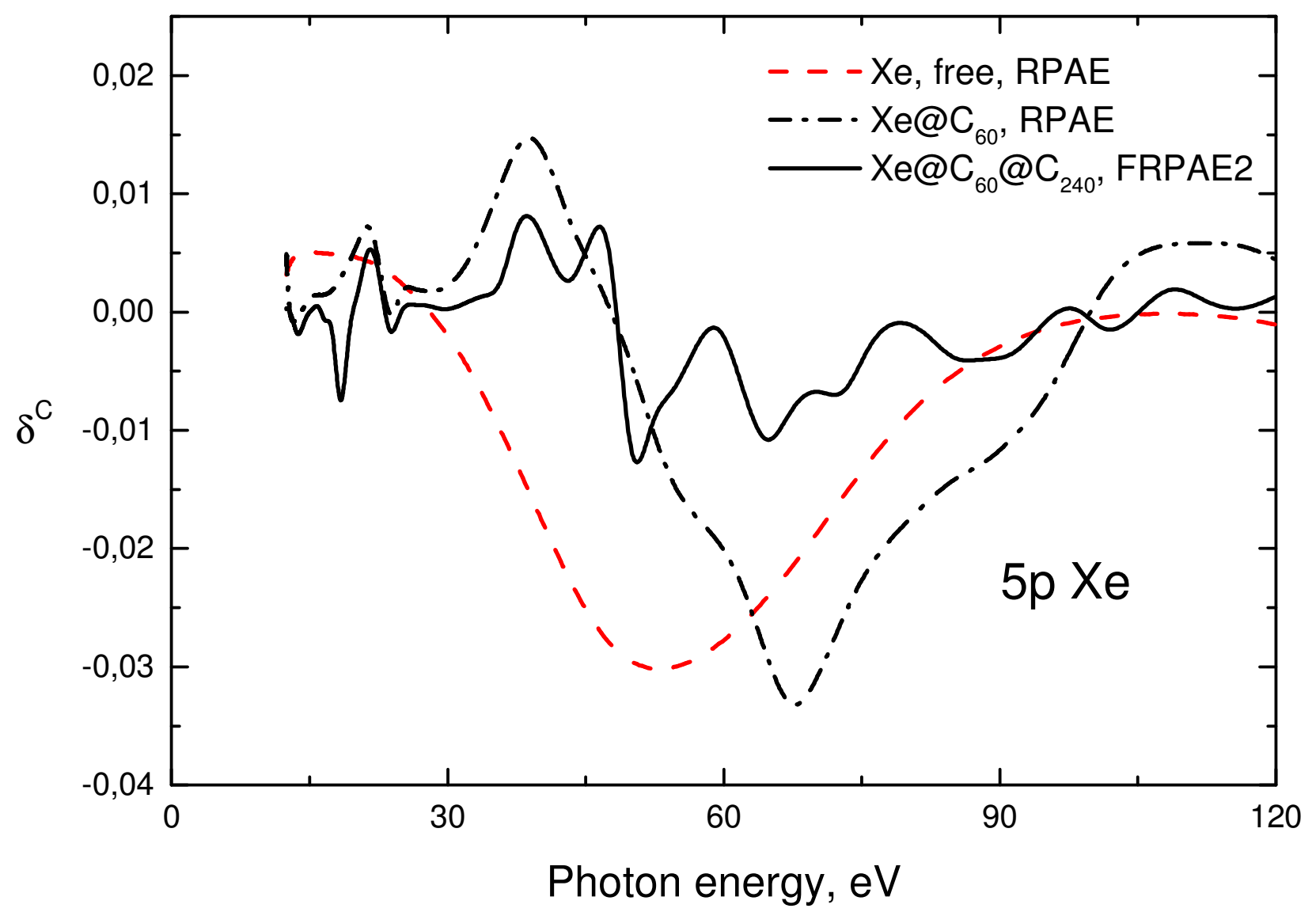

Fig.18. Non-dipole angular anisotropy parameter $\delta_{5p}^{C}(\omega)$ of 5p electrons in Xe, $Xe@C_{60}$ and $Xe@C_{60}@C_{240}$ with account of reflection factors F.

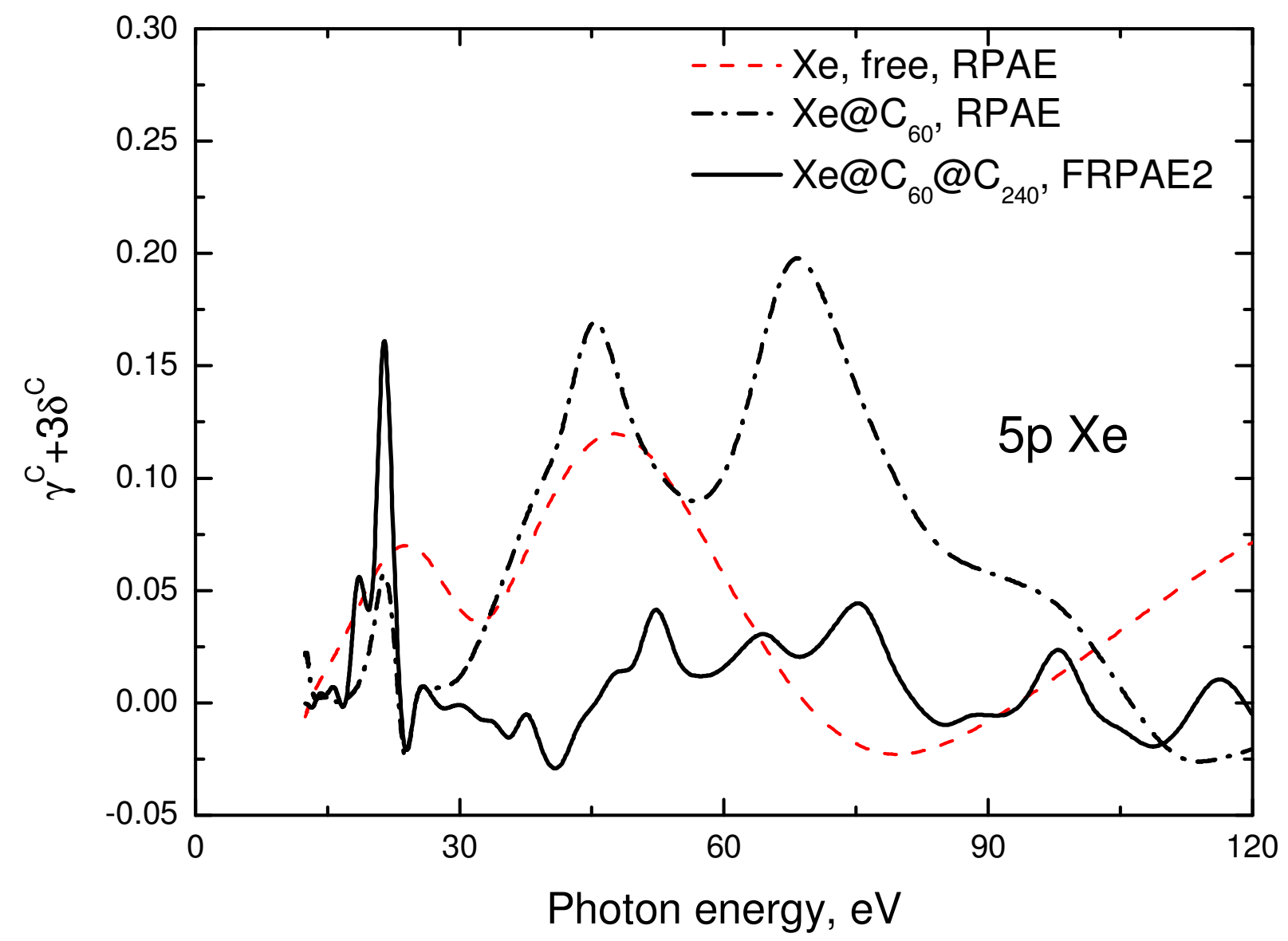

Fig.19. Combination of non-dipole angular anisotropy parameters $\gamma^C_{3p}(\omega)+3\delta^C_{3p}(\omega)$ for 5p electrons in Xe, Xe@$C_{60}$ and Xe@$C_{60}$@$C_{240}$ with account of reflection factors F.

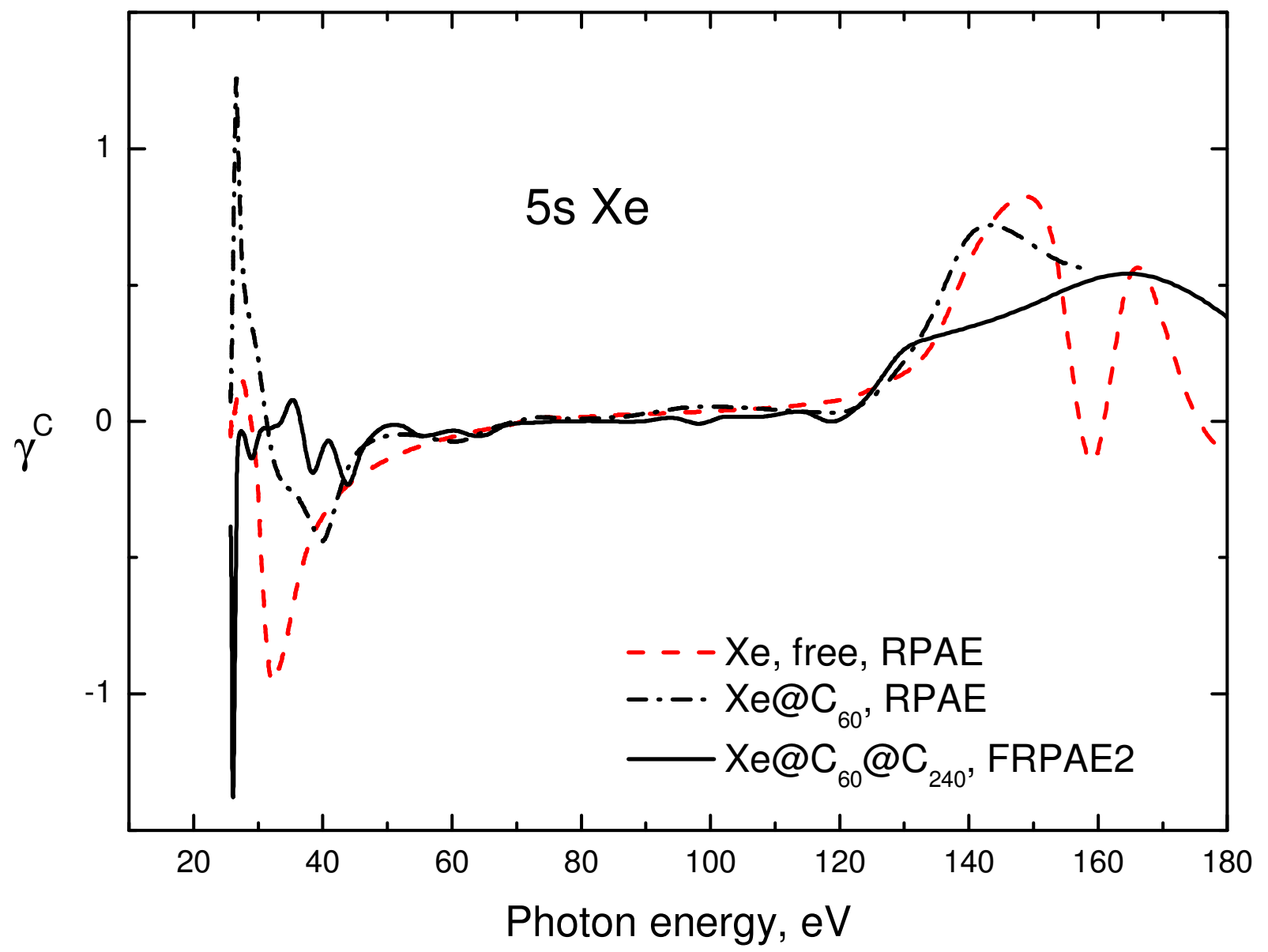

Fig.20. Non-dipole angular anisotropy parameter $\gamma^C_{5s}(\omega)$ of 5s electrons in Xe, Xe@$C_{60}$ and Xe@$C_{60}$@$C_{240}$ with account of reflection factors F.

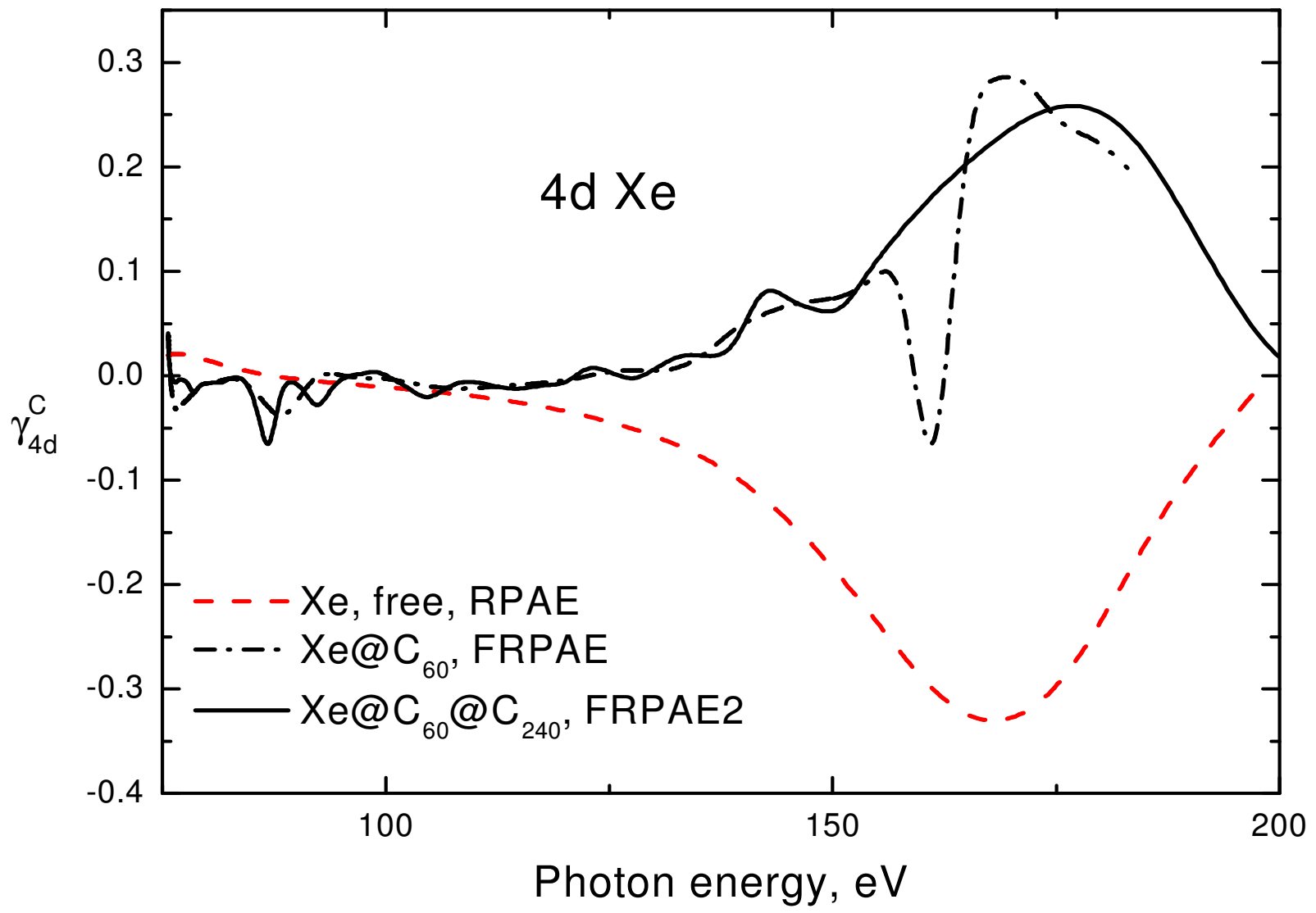

Fig.21. Non-dipole angular anisotropy parameter $\gamma^C_{4d}(\omega)$ of 4d electrons in Xe, Xe@$C_{60}$ and Xe@$C_{60}$@$C_{240}$ with account of reflection factors F.

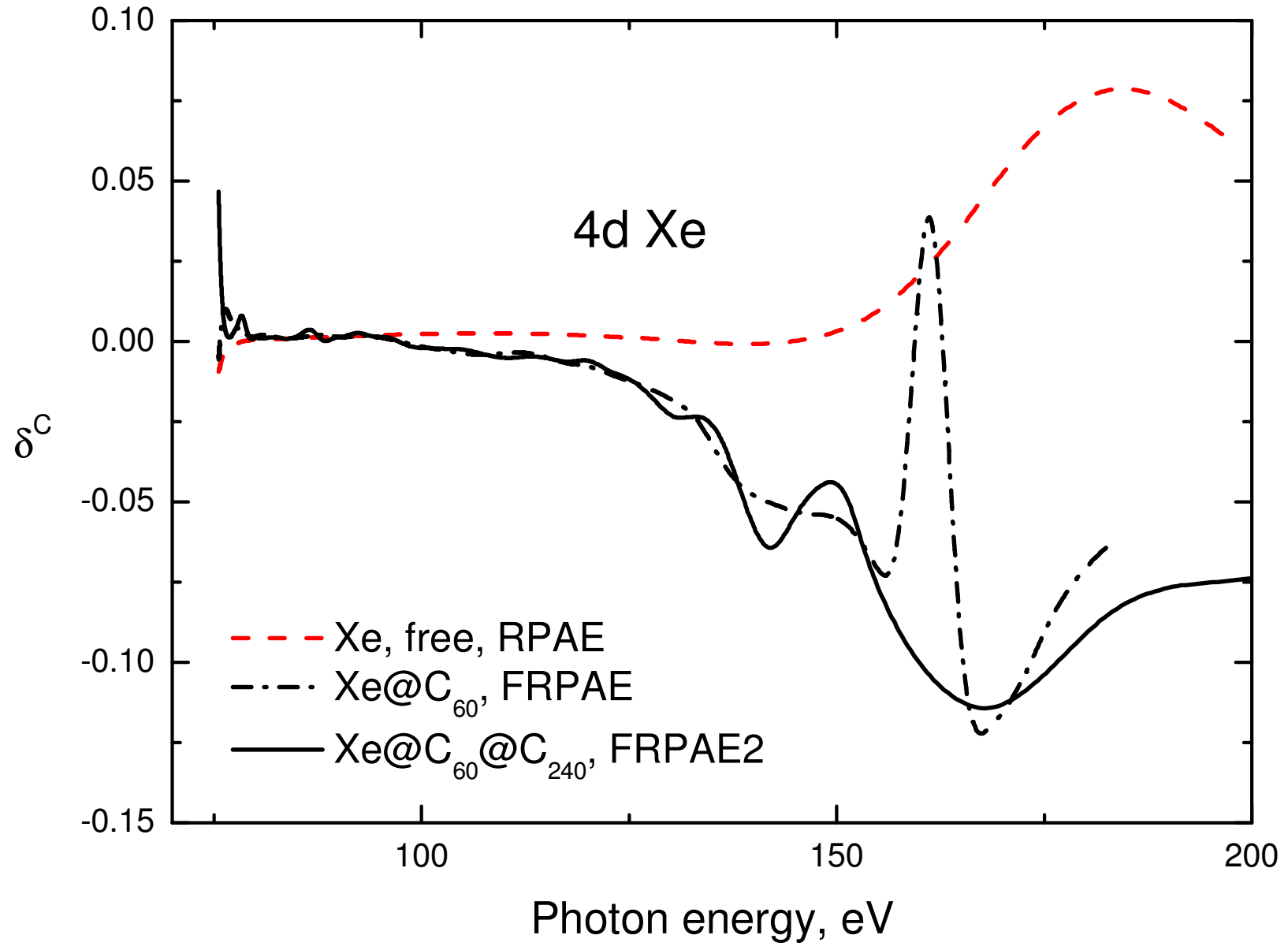

Fig.22. Non-dipole angular anisotropy parameter $\delta^C_{4d}(\omega)$ of 4d electrons in Xe, Xe@$C_{60}$ and Xe@$C_{60}$@$C_{240}$ with account of reflection factors F.

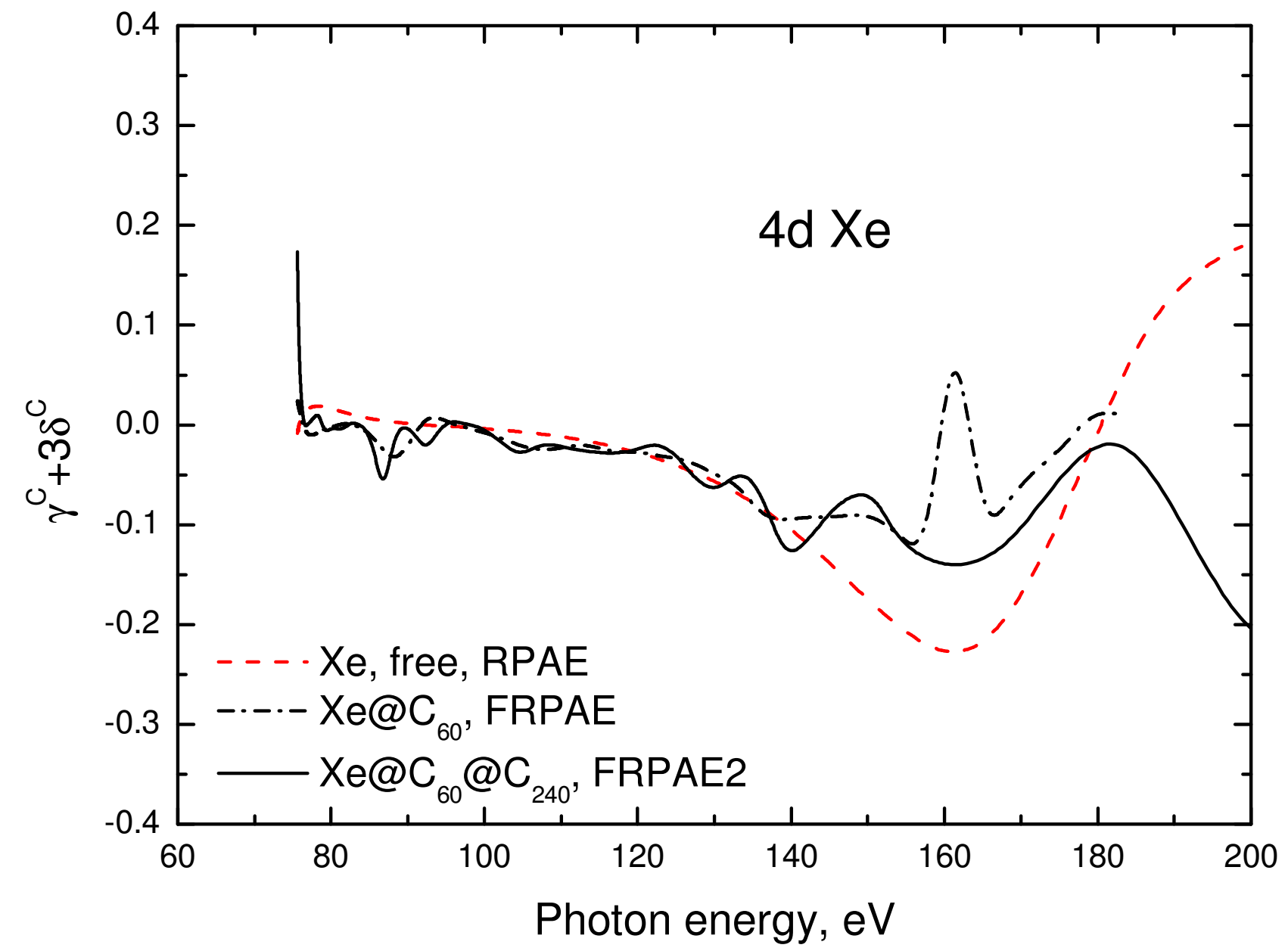

Fig.23. Combination of non-dipole angular anisotropy parameters $\gamma_{4d}^{C}(\omega)+3\delta_{4d}^{C}(\omega)$ for 4d in Xe, Xe@$C_{60}$ and Xe@$C_{60}$@$C_{240}$ with account of reflection factors F.